\documentclass[lettersize,journal,twoside]{IEEEtran}

\usepackage{booktabs}
\usepackage{multicol}
\usepackage{multirow}
\usepackage{graphicx}
\usepackage[hidelinks]{hyperref}
\usepackage{booktabs}
\usepackage{amssymb}
\usepackage{amsmath,amsfonts}
\usepackage{wasysym}
\usepackage{color}
\usepackage{url}
\usepackage{cite}
\usepackage{graphicx}
\begin{document}

	\title{Decentralized Federated Learning: A Survey on Security and Privacy}
	
	\author{Ehsan~Hallaji,~\IEEEmembership{Graduate Student Member,~IEEE,}
		Roozbeh~Razavi-Far,~\IEEEmembership{Senior Member,~IEEE,}
		Mehrdad~Saif,~\IEEEmembership{Fellow,~IEEE,}
		Boyu~Wang,~\IEEEmembership{Member,~IEEE,}
		Qiang~Yang,~\IEEEmembership{Fellow,~IEEE}
        \thanks{This work was supported by the Natural Sciences and Engineering Research Council of Canada (NSERC).}
		\thanks{Ehsan Hallaji and Mehrdad Saif are with the Department of Electrical and Computer Engineering, University of Windsor, Windsor, ON N9B 3P4, Canada. E-mail: hallaji@uwindsor.ca, msaif@uwindsor.ca.\protect}
		\thanks{Roozbeh Razavi-Far is with the Faculty of Computer Science, University of New Brunswick, Fredericton, NB E3B 5A3, Canada, and also with the Department of Electrical and Computer Engineering, University of Windsor, Windsor, ON N9B 3P4, Canada. E-mail: roozbeh.razavi-far@unb.ca.\protect}
		\thanks{Boyu Wang is with the Department of Computer Science and the Brain Mind Institute, University of Western Ontario, London, ON N6A 5B7, Canada. E-mail: bwang@csd.uwo.ca.\protect}
		\thanks{Qiang Yang is with the Department of Computer Science and Engineering, Hong Kong University of Science and Technology, Clear Water Bay, Kowloon, Hong Kong, and also with the Department of AI, WeBank, Gaungdong Province, China. E-mail: qyang@cse.ust.hk.}
		
	}

\markboth{IEEE Transactions on Big Data, January 2024}%
{Hallaji \MakeLowercase{et al.}: Decentralized Federated Learning: A Survey on Security and Privacy}

\maketitle

\begin{abstract}
Federated learning has been rapidly evolving and gaining popularity in recent years due to its privacy-preserving features, among other advantages. Nevertheless, the exchange of model updates and gradients in this architecture provides new attack surfaces for malicious users of the network which may jeopardize the model performance and user and data privacy. For this reason, one of the main motivations for decentralized federated learning is to eliminate server-related threats by removing the server from the network and compensating for it through technologies such as blockchain. However, this advantage comes at the cost of challenging the system with new privacy threats. Thus, performing a thorough security analysis in this new paradigm is necessary. This survey studies possible variations of threats and adversaries in decentralized federated learning and overviews the potential defense mechanisms. Trustability and verifiability of decentralized federated learning are also considered in this study.
\end{abstract}


\section{Introduction}\label{sec:introduction}

\IEEEPARstart{W}{ith} 
the widespread use of machine learning over the recent years, new concerns have been raised regarding user and data privacy. The data-driven nature of these intelligent models necessitates gathering users' data to constantly improve and maintain the operating statistical model. This issue becomes more problematic in large-scale distributed systems with millions of users such as mobile networks. 

Aside from privacy issues, in large-scale systems, communicating user data may pose an overhead to the network. This is while information technology and intelligent devices are evolving at a rapid pace, and in the wake of it, there is an explosion of data at the edge of the network. Given the potential benefits of this data collection process for improving their model, organizations tend to make the best use of it for knowledge extraction with minimal waste of data. Thus, modern intelligent systems struggle to find the optimal trade-off between user privacy and service quality.

\begin{figure*}[t]
	\centering
	\includegraphics[trim={0.5cm 0.7cm 0.5cm 0.5cm},clip,width=\textwidth]{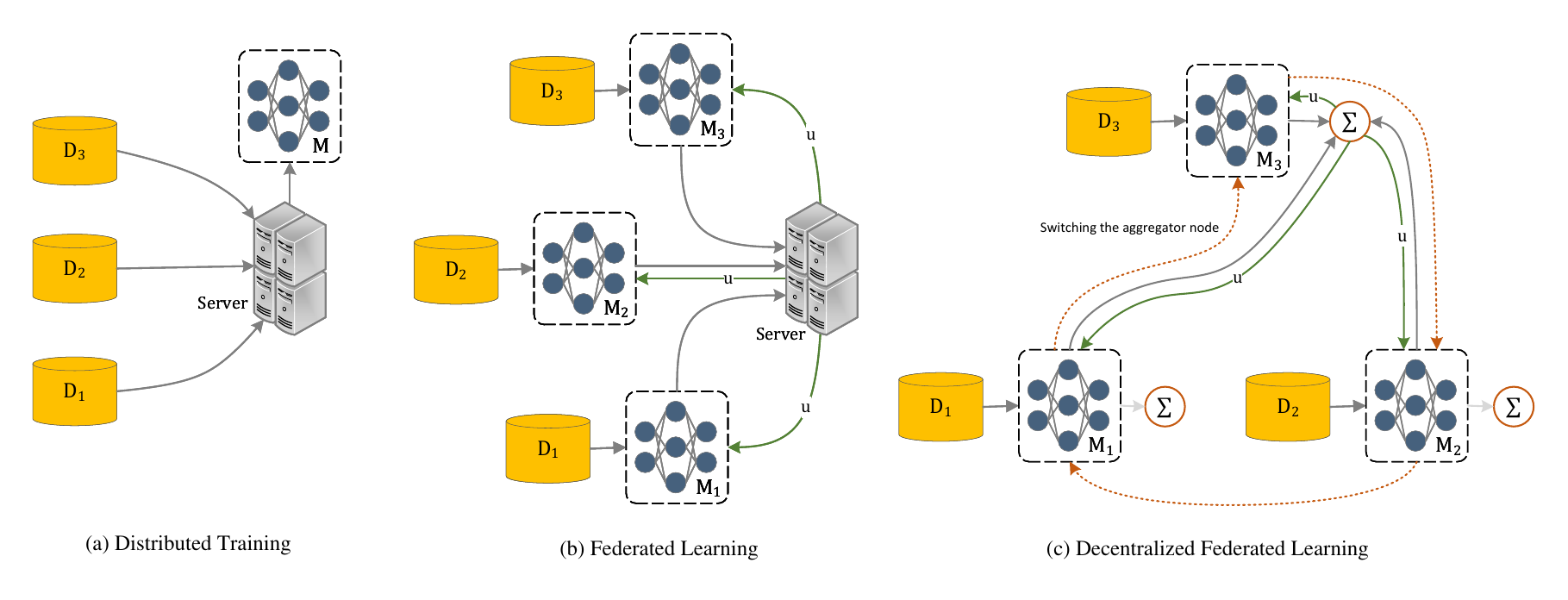}
	\caption{Workflow of distributed training, FL, and DFL with three nodes. $D_i$, $M_i$, and $u$ denote client data, local model parameters, and the generated update by the server. In this example, three clients are shown in the picture (i.e., $1\leq i\leq 3$).}
	\label{fig:workflow}
\end{figure*}

Inspired by recent breakthroughs in distributed optimization \cite{DBLP:journals/corr/KonecnyMRR16}, Federated Learning (FL) has proposed as a potential solution to resolve the aforementioned challenges \cite{pmlr-v54-mcmahan17a}. In contrast to distributed machine learning which is shown in Fig. \ref{fig:workflow}(a), FL proposes training local models at the edges of the network and then sharing the model parameters to a central server that aggregates the received information and then updates all the client models. 

Aside from its use in distributed training of Deep Learning (DL) models, FL offers additional advantages. Firstly, by communicating the local model parameters rather than the user's data, the central model is trained on decentralized data, which no longer jeopardizes data privacy. Secondly, the communication is extensively reduced for large-scale systems, as the volume of model parameters is often smaller than the training data itself. The general scheme of FL workflow is illustrated in Fig. \ref{fig:workflow}(b). As shown in this figure, for client $i$, a model $M_i$ is constructed locally based on the local user data $D_i$. The model parameters $M_i$ are then sent to a server that aggregates all models in the form of $\sum (M_1,M_2,M_3) = u$, where $\sum(\cdot)$ and $u$ are the aggregation function and the generated update, respectively. Finally, $u$ is sent back to the clients to update local models. 

Despite the breakthrough made by FL, the proposed architecture was not flawless and demanded further research endeavors on the topic. To begin with, although the user data is not being shared in the network, communicating local parameters is still vulnerable to sniffing attacks which will lead to stealing model parameters for launching an inference attack to extract sensitive information from the local training data at the edges of the networks \cite{9048613}. In addition, the trustability of the central server is sometimes questionable in environments such as wireless networks.  Moreover, having a Single Point of Failure (SPF) in the system is not ideal since a compromised server can affect the whole network \cite{9019859}. While various works have been dedicated to designing defense mechanisms for FL, it was proposed that decentralizing the FL architecture will eliminate part of these security challenges \cite{9292450}.

Relying on Peer-to-Peer (P2P) communications, Decentralized Federated Learning (DFL) improves both the reliability and scalability of FL by eliminating SPF and enhancing the communication efficiency \cite{9084352, DBLP:journals/corr/abs-1912-04977}. The process of local update preparation is similar to that of FL; however, exchanging model parameters and model aggregation is mostly undertaken through P2P communication or blockchain technology. For instance, a generic diagram of a DFL workflow is shown in Fig. \ref{fig:workflow}(c), where a client is assigned with the aggregator role in each round (i.e., shown with dotted lines) to apply the aggregating function $\sum(M_1,M_2,M_3)=u$ on the received parameters and update all nodes with $u$. The protocol for selecting the aggregating node and the $\sum(\cdot)$ function varies among different DFL architectures. Integration with blockchain brings about additional advantages including traceability and immutability. Despite the security and efficiency features of DFL, this architecture is not flawless. For instance, incorporating blockchain into FL may come at the cost of making the system defenseless against blockchain-related security adversaries. The trustability of the DFL participants is another issue of concern that requires further research.

\begin{table}
	\centering
	\caption{Abbreviations used in this article. The list is sorted alphabetically.}
	\begin{tabular}{ll}
		\toprule
		Abbreviation & Definition \\
		\midrule
		BAFFLE & Blockchain-based aggregator free FL\\
		BEAS & Blockchain enabled asynchronous and secure FL\\
		BFLC & Blockchain-based FL with committee consensus\\
		BindaaS & Blockchain-based deep learning as-a-service\\
		DDoS & Distributed denial of service \\
		DFL & Decentralized federated learning\\
		DL & Deep learning\\
		DoS & Denial of service \\
		DP & Differential privacy\\
            $f$-DP & Functional differential privacy\\
		FL & Federated learning\\
		FTL & Federated transfer learning \\
		GAN & Generative adversarial network\\
            HE & Homomorphic encryption\\
		HFL & Horizontal federated learning \\ 
		IoT & Internet of things \\ 
		P2P & Peer-to-peer \\
		PoS & Proof of state\\
		PoW & Proof of work\\
		TEE & Trusted Executive Environment\\
		TL & Transfer learning\\
		VFL & Vertical federated learning\\
		SC & Smart contract \\
		SGD & Stochastic gradient decent \\ 
		SL & Swarm learning\\
		SMC & Secure multiparty computation\\
		SPF & Single point of failure\\
		\bottomrule
	\end{tabular}
	\label{tab:abrv}
\end{table}

Currently, the literature lacks a survey that primarily studies the security and privacy of DFL. Available reviews on DFL generally study trending approaches in DFL with an emphasis on the application in the Internet of Things (IoT) \cite{9403374, li_blockchain_2021, 9441499, 9540163, DBLP:journals/corr/abs-2110-02182, ALI2021102355}. On the other hand, surveys on security and privacy of centralized FL \cite{DBLP:journals/corr/abs-2003-02133,9308910,DBLP:journals/corr/abs-2012-06337,9411833,MOTHUKURI2021619, liu_threats_2022} do not discuss the applicability of their analysis and findings to DFL. Among the limited number of surveys on DFL, \cite{cite-key} studies security issues in centralized FL that can be addressed through blockchain integration. However, the literature still lacks a thorough security analysis of threats in DFL and its potential defense mechanisms. Hence, it is worthwhile to perform a comprehensive survey on the new attack surface created on DFL, and defense mechanisms that can be directly used or adapted for DFL. Toward this goal, this survey makes the following contributions:
\begin{itemize}
	\item State-of-the-art DFL methods are reviewed in terms of security robustness and employed technologies.
	\item Potential threats to DFL systems are identified and explained.
	\item Defense mechanisms that can safeguard DFL systems against attacks are studied and analyzed.
	\item The effect of blockchain integration on the security and privacy of DFL is studied.
	\item The connection between verifiable FL and DFL is studied from a security standpoint.
	\item Undiscovered domains and demanding research directions in enhancing DFL security and privacy are identified and introduced.
\end{itemize}

Reviewed works in this survey are collected from Scopus and Google Scholar searches. The keywords used for the search are FL, DFL, security, privacy, attack, defense, and blockchain. After the initial search, irrelevant papers to the topic of this survey were excluded. Furthermore, the search was limited to works after 2015, and we consider published or in-press research works, as well as arXiv preprints.

Towards this, we outline the background for this survey in Section 2. Preliminaries of DFL are reviewed in Section 3. Section 4 studies possible threats to DFL. Section 5 discusses the potential defense mechanisms in DFL. Section 6 elaborates on the verifiability of DFL. Future research directions are discussed in Section 7. Finally, the survey is concluded in Section 8. This survey contains several abbreviations to make discussions more concise, which are listed in Table \ref{tab:abrv}.

\section{Background}
\label{sec:back}
DFL is a paradigm that often encompasses both FL and blockchain technologies, although it is important to note that DFL can exist independently of blockchain. To better comprehend the concept of DFL, it is crucial to have a clear understanding of the fundamentals of FL and blockchain. Thus, the following section provides a concise overview of these concepts.

\subsection{Federated Learning}
FL was proposed to reduce the risks of data ownership in collaborative training of DL models. Prior to the introduction of FL, collaborative training required immense loads of data exchange between participants of a distributed machine learning framework in which the clients constantly send local training data to a central server. The distributed model is then trained on the accumulated training data gathered from across the network. Nevertheless, communicating user data and overburdening the network with excessive communication load raised new concerns that motivated the invention of FL \cite{pmlr-v54-mcmahan17a}. In simple words, rather than sharing the training of the local data with the server, the client in FL locally trains a model of the same structure on the user data and sends the obtained model parameters to the server once convergence is achieved. The server aggregates these parameters (e.g., neural network weights) and updates all the client models with the aggregated parameters. As this cycle continues, in each round, clients initialize their local model parameters with the received update from the server before starting the local training and issuing an update to the server. It is worth mentioning that the collaboration schemes between clients at the edge of the network have been a topic of interest in the domain of multi-agent systems even before the birth of FL \cite{10.1007/3-540-45448-9_10}. For instance, client-server collaboration in FL somehow resembles the umbrella system in which all agents communicate with a server \cite{Hayden1999ArchitecturalDP}.

The aforementioned process in FL, however, assumes the homogeneity of data among all clients, which may not always be the case. This leads to the evolution of FL into a more advanced structure for dealing with heterogeneous data and comes with additional security perks.  Specifically, \cite{10.1145/3298981} categorizes techniques in this field into three groups: 1) Horizontal FL (HFL)  \cite{1316832}, 2) Vertical FL (VFL) \cite{DBLP:journals/corr/abs-1711-10677, 10.1145/3298981}, and 3) Federated Transfer Learning (FTL) \cite{9076003, 10.1145/3298981}. The difference between these groups is in the nature of the training data in the FL systems. HFL which is basically defined based on the initial FL model in \cite{1316832}, assumes a diverse range of samples across all local training sets that all fall under a similar feature space. VFL on the other hand, considers a common sample space among participants who differ in the feature space. FTL addresses the minimal overlap between both sample and feature space by resorting to transfer learning for transferring knowledge between participants. 

It is worth mentioning that this categorization has primarily focused on centralized FL. Therefore, the focus of this survey is primarily on DFL which differs from the centralized FL. The central server in the FL systems is in charge of verifying the network clients, aggregating the local parameters, and broadcasting an update. With the absence of servers in the DFL structure, these tasks must be performed using the client networks. The current literature usually devises the concept of blockchain to carry out the mentioned tasks.

Even though the initial FL model was proposed on the basis of DL, decision tree models can also be employed and adapted to FL frameworks \cite{10.14778/3407790.3407811, 9440789, Li_Wen_He_2020}. Threats and defense mechanisms that will be later reviewed in this work generally apply to both DL and tree-based models.

\subsection{Blockchain}
Blockchain is designed primarily with the objective of making the data exchange immutable and traceable over a P2P network \cite{nakamoto2008bitcoin}. Similar to FL, in a blockchain network data is decentralized, and each user will transfer data using a block. This consists of the data, a hash acting as a unique identifier for the block, and a previous hash pointing to the previous block in the chain. While these hashes make the data traceable in the chain, they also complicate tampering with data, as changing the data in an existing block will also alter the associated hash which creates a discrepancy in the chain. 

\begin{figure*}[t]
    \centering
    \includegraphics[trim={0.7cm 0.6cm 0.7cm 0.6cm},clip,width=0.95\textwidth]{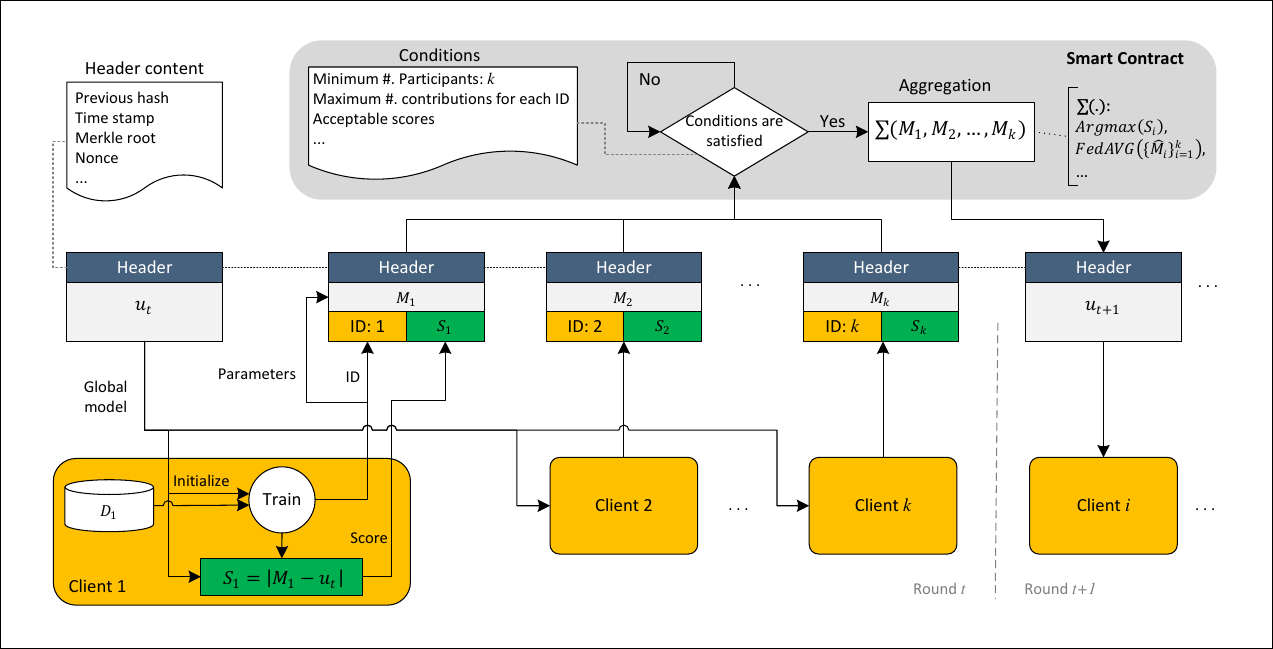}
    \caption{Generic process of blockchain-based DFL. SC can use any choice of aggregation mechanism such as selecting the update with the highest score or averaging all received model parameters. $S_i$ indicates the calculated score for the estimated model parameters for client $i$.}
    \label{fig:blockchain}
\end{figure*}

A key idea behind the blockchain is to prevent a certain participant from controlling the network and provide all network members with an equal chance to verify and control the transactions \cite{Nguyen2020, 9403374}. Nonetheless, finding the optimal solution to ensure fairness in assigning the evaluators is still an open problem and as a result, many consensus mechanisms have been proposed over time. Perhaps, the most common consensus mechanisms are Proof of Work (PoW) and Proof of State (PoS), albeit they are not necessarily the best. In PoW, the participants (also called miners) need to solve a time-consuming problem and anyone who achieves the answer quicker will get a chance to contribute to the chain. As an alternative to PoW, PoS randomly selects the evaluators with the aim of improving the scalability.

Based on the current literature, Blockchain approaches can fall into four major categories, namely 1) public, 2) private, 3) consortium, and 4) hybrid blockchain (i.e., we refer to this as DFL)  \cite{Wang2019, 9403374, WANG201910}. These variations are briefly explained in the following.

\begin{enumerate}
	\item \textbf{Public:} Blockchain is built in a permissionless network that allows anyone to join the network and participate in the consensus process \cite{nakamoto2008bitcoin}.
	\item \textbf{Private:} Participants can join the network through a received authorization \cite{ethereum, hyperledger}. While permissions are granted in a centralized manner, the consensus mechanisms are decentralized \cite{Buterin2013, Androulaki2018}.
	\item \textbf{Consortium:} Blockchain is semi-decentralized and the network access is managed by more than one entity \cite{Kang2017, Li2018}.
	\item \textbf{Hybrid:} Block construction is carried out within a private network and the storage is done through a public network \cite{Watanabe2016, Cui2020}.
\end{enumerate}

\section{Decentralized Federated Learning}

\begin{figure*}[t]
	\centering
	\includegraphics[width=\textwidth]{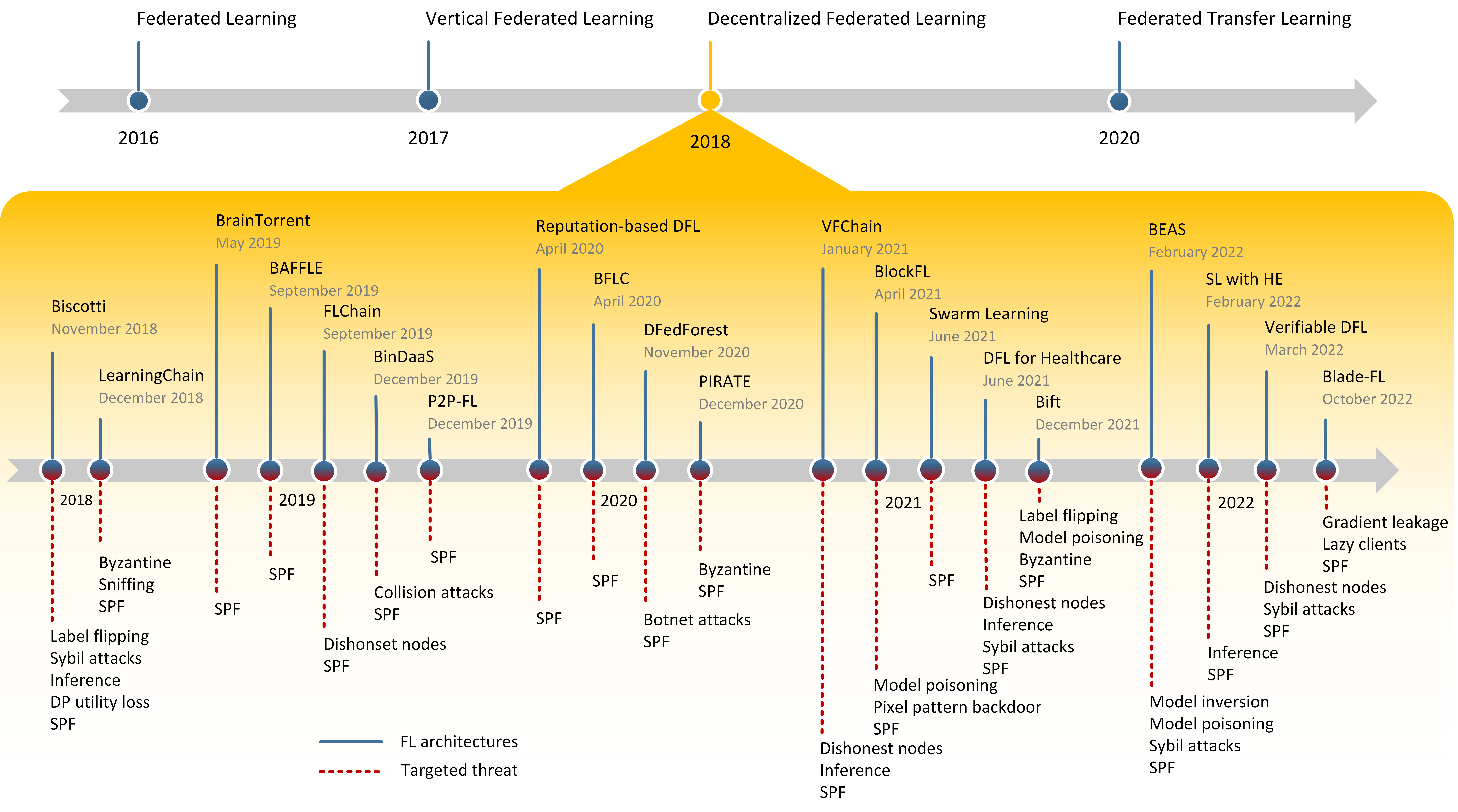}
	\caption{Advances in decentralized federated learning in time. The order of the methods is set based on the first version that became available online (e.g., pre-prints).}
	\label{fig:diagram}
\end{figure*}

\label{sec:DFL}
As mentioned, the main flaw with the centralized FL is the dependency of the entire federation on a central server whose privacy and performance directly affect all the clients in the system \cite{pmlr-v54-vanhaesebrouck17a, DBLP:journals/corr/abs-2107-12048,NIPS2017_f7552665}. In other words, a compromised server will jeopardize the whole federated learning system \cite{10.1145/3485875}. To eliminate the dependency of the whole network on a single node and also enhance the communication efficiency \cite{DBLP:journals/corr/abs-1908-07782}, DFL resorts to P2P communications to bypass the need for a central server. To do so, model aggregation and participant verification should be carried out in a serverless fashion. 

As explained in Section \ref{sec:back}, DFL often uses blockchain to facilitate inter-node communications. While this combination is non-absolute, blockchain can effectively facilitate communication by treating model updates as data within a block and sharing the block with respect to a consensus mechanism. In this process, local model construction is similar to that of FL, and prepared updates will be in the form of a data block. The prepared block will be then shared with the rest of the nodes for consensus and addition to the blockchain. The aggregation mechanisms of DFL work on the basis of decentralized Stochastic Gradient Decent (SGD), which makes use of gossip averaging \cite{NIPS2017_a74c3bae}.

DFL models that use blockchain have distinct architectures. For instance, the block structure, information included in the headers, aggregation, and consensus mechanisms can all be different in each architecture. However, a generic scheme is shown in Fig. \ref{fig:blockchain} that resembles the most common practices in designing a blockchain-based DFL. In this example, the blockchain is initialized using a single block containing the starting global model. Each update is added as a new block that is linked to the previous block in the chain using a header. On the client's side, the local model is initialized using the global model $u_t$ and trained on the $i$-th client data $D_i$. Once the trained model $M_i$ is obtained, a score $S_i$ is estimated w.r.t. the difference between the global model and the estimated local model. Nevertheless, not all blockchain-based DFL models necessarily use a scoring system. The client then uploads data to the blockchain by creating a block containing a header, trained model, estimated score, and the uploader identifier (ID). In some architectures, clients are directly communicating with a set of miners to obtain the Merkle root of the data before uploading the block. It is also worthwhile to mention that the block size is often limited to a fixed size. If the model size is larger than the block size, the uploader has to upload a set of serialized blocks to the blockchain \cite{9284684}. The SC on this blockchain uses a set of rules to determine when the global model should be updated. For instance, a minimum number of unique clients $k$ are required to participate in updating $u_t$. Furthermore, considering that each client is allowed to serialize its updates into multiple blocks, a limit is set to define the maximum number of contributions by each client in each round. The aggregation mechanism varies in different DFL architectures. While the majority of DFL models use an averaging function $\sum(\cdot)$ similar to that of federated averaging (FedAVG) \cite{pmlr-v54-mcmahan17a}, other frameworks use different approaches such as selecting the update with the highest score \cite{9284684}.

While the initial DFL model \cite{DFL, DBLP:journals/corr/abs-1901-11173} was proposed with the aim of security enhancement and failure robustness of FL using Bayesian beliefs of one-hop neighbors, several research studies extend their work by improving the overall efficiency and security. Fig. \ref{fig:diagram} illustrates the historical evolution of DFL since it was first proposed in 2018. BAFFLE \cite{9284684} makes use of Smart Contracts (SC) to facilitate storing the FL model and clients' states.  Model updates and aggregation is also carried out by resorting to SC. BAFFLE takes the computational states into account in order to preserve fairness among clients. Another approach, Blade-FL \cite{9664296}, uses clients for both mining and training via gossip learning. A novel consensus mechanism was proposed in \cite{9293091} to enhance the efficiency and privacy of the system. The DFL framework proposed based on this committee consensus mechanism is called BFLC. The Biscotti technique \cite{9292450} integrates DFL with Differential Privacy (DP) to safeguard DFL against a range of attacks. Focused on healthcare applications, \cite{8943171} introduces the BindaaS method, which integrates DL as a service and lattices-based cryptography. This enables BinDaaS to employ an authentication phase that enhances DFL attack resilience. Another DP-based DFL approach called LearningChain is proposed with the aim of combating Byzantine attacks \cite{8622598}. To do this, LearningChain utilizes a new aggregation mechanism for decentralized SGD. Reference \cite{9210138} designs the PIRATE framework based on the sharding approach in blockchain which helps in securing the aggregation process. BEAS \cite{DBLP:journals/corr/abs-2202-02817} employs DP, gradient pruning, and anomaly detection to protect DFL against poisoning attacks. BEAS is also adapted for heterogeneous data to ensure proper convergence of the global model. BrainTorrent \cite{roy2019braintorrent} is a DFL method that employs version control and arbitrarily assigns a participant to act as the server. A reputation management scheme is used in \cite{8994206} to select updates from reliable workers in the blockchain which reduces the probability of successful poisoning attacks in the DFL system. BlockFLA \cite{10.1145/3422337.3447837}, uses smart contracts to detect and penalize malicious nodes by means of monetary penalty. BlockFLA requires uploading hashed updates into separate private and public blockchains. Smart contracts is used for aggregating gradients whereas hashed updates are used for evaluating the updates through recovering the parameters to detect any mismatch. DFedForest is a tree-based DFL framework that uses bagging for tree construction \cite{9284805}. This method devises a private test dataset for evaluating the uploaded parameters as malicious updates are expected to result in anomalous outputs on this data. Bift, \cite{9650783}, designs a Proof of FL by combining PoW with FederatedAVG \cite{pmlr-v54-mcmahan17a} and FederatedSGD \cite{45187} aggregation schemes to reach secure DFL with low communication overhead. A verifiable version of DFL \cite{FANG20221} incorporates private key sharing and gradient masking into the consensus mechanism to defend against malicious miners and dropouts. Swarm Learning (SL) also integrates FL and blockchain to distributively train a model for diagnosing a number of diseases while safeguarding the patient's data privacy and security \cite{warnat-herresthal_swarm_2021}. Security and privacy features of SL are limited to those of blockchain and the P2P DFL design. However, there is an extension of SL that couples it with HE is also presented in \cite{10.1007/978-3-030-95391-1_32} to secure SL communications against privacy attacks. Various aspects of SL such as fault tolerance, scalability, and fairness are studied in \cite{DBLP:journals/corr/abs-2201-05286}, and the results indicate an overall improvement over centralized FL. 

Despite the security advantages of blockchain integration with DFL, communication delay and resource consumption issues often become problematic in this domain \cite{Nguyen2020}. As mentioned in Section \ref{sec:back}, the process of model aggregation and achieving a consensus often requires sufficient computational power at the edge of the network which may not always be practical (e.g., mobile networks). This becomes more critical for consensus mechanisms such as PoW. As a result, the model aggregation process may be usually delayed due to the potential computational constraints.

\section{Threats to Decentralized Federated Learning}
\label{sec:threats}
\begin{figure*}[t]
	\centering
	\includegraphics[width = \textwidth, trim = {0.9cm 0.9cm 0.9cm 0.9cm},clip,page=4]{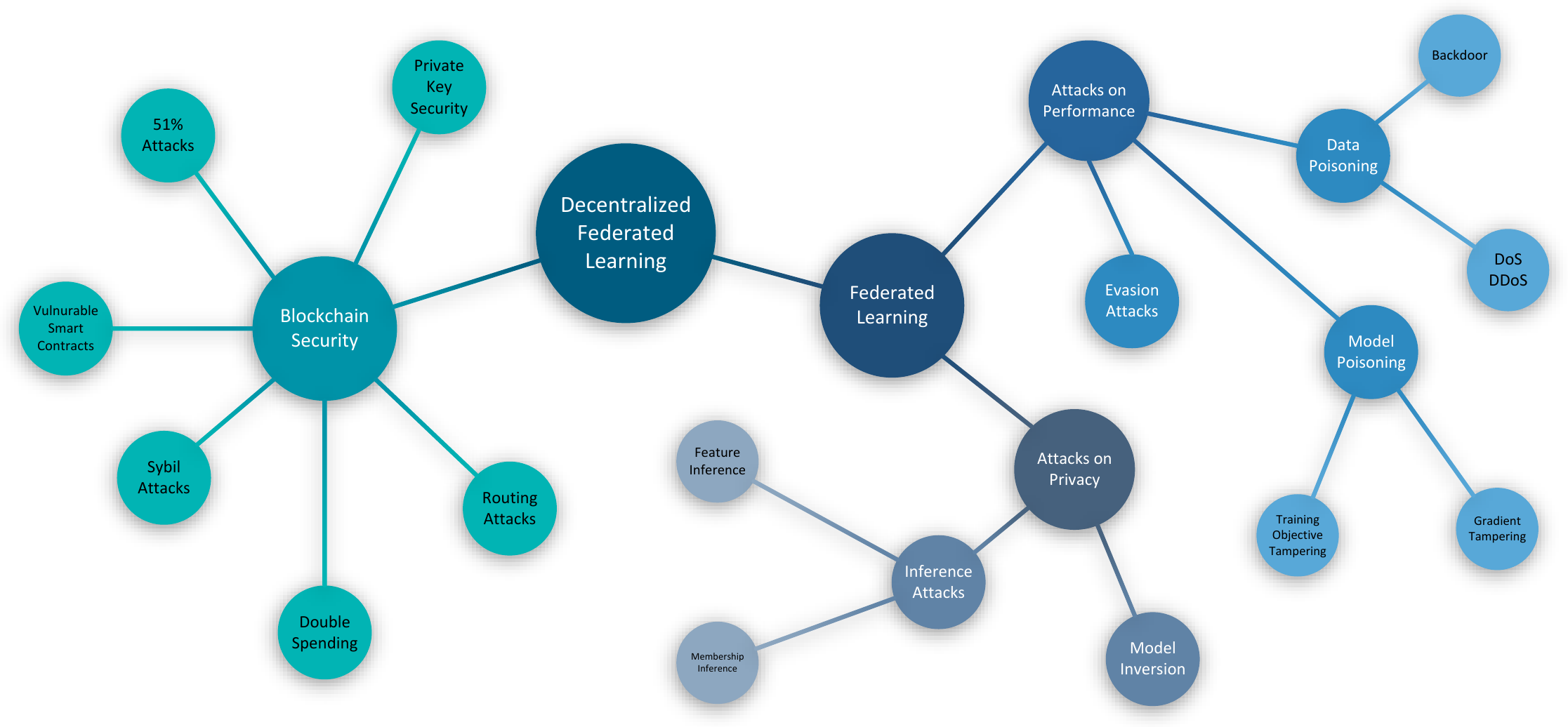}
	\caption{Overview of security and privacy threats on decentralized federated learning. Threats are generally related to privacy, model robustness, or blockchain bottleneck of DFL.}
	\label{fig:taxonomy}
\end{figure*}

As shown in Fig. \ref{fig:taxonomy}, the security and performance of DFL are directly related to those of FL and blockchain. Generally, threats to DFL target privacy or the performance of the global model. In centralized FL, all the possible threats involve the server at some point as the server is the core of the system for aggregation, communication, and validation. As a result, the trustworthiness of the server is of great importance in FL as a malicious server can easily attack all clients. On the other hand, attacking the server and causing a malfunction can also disable the entire system. Table \ref{tab:tab1} lists sources of threats for common attacks on FL. However, in DFL, all possible threats will be traced back to the clients or miners. Table \ref{tab:analysis} indicates attack models that are considered in the design of the aforementioned DFL techniques.

\begin{table}[t]
	\centering
	\setlength{\tabcolsep}{4pt}
	\caption{Identification of sources of attacks in FL systems.}
	\begin{tabular}{ll}
		\toprule
		Attacks & Source of Attack \\
		\midrule
		Data poisoning & Malicious client \\
		Gradient manipulation & Malicious client\\
		Backdoor attack & Malicious client and malicious server\\
		Evasion attack & Malicious client and model deployment\\
		Non-robust aggregation & Aggregation algorithm\\
		Training rule manipulation & Malicious client\\
		Inference attacks & Malicious server and communication\\
		GAN reconstruction & Malicious server and communication\\
		Free-riding attack & Malicious client\\
		Man-in-the-middle attack & Communication\\
		\bottomrule
	\end{tabular}
	\label{tab:tab1}
\end{table}

Malicious clients can follow different attack models to jeopardize the performance or privacy of the system \cite{DBLP:journals/corr/abs-2012-06337}. For instance, an adversary can be either semi-honest or aggressive. In the semi-honest model, the objective is to infer confidential information while complying with the DFL protocol. This phase can be also used as a reconnaissance step prior to launching an aggressive attack where the aim is to degrade the system's performance. Using this combination, one can first learn the global model parameters using a semi-honest approach and then actively forge malicious updates in order to insert a backdoor in the global model. Due to the immutability of blockchain, hackers cannot directly manipulate the training data in a block. Nevertheless, they can reach their objective by sending the gradients that lead to incorrect predictions. To do so, gradients can be forged targeted or untargeted. As the name implies, the latter aims at degrading the performance of all classes in general whereas the targeted scheme only attacks a certain class without tampering with the rest of the model \cite{DBLP:journals/corr/abs-2012-06337}.

\subsection{Attacks on Performance}
The performance of the DFL model is mostly targeted by means of poisoning attacks. In addition, there are certain blockchain-based attacks that also apply to DFL. An overview of these attacks for DFL is given in the following.

\subsubsection{Data Poisoning}
Data poisoning techniques in FL and DFL could be somewhat different. In centralized FL, data poisoning is referred to as the process of corrupting the local training data via backdoor injection or introducing label noise which in turn produces deviating gradients and misleads the FL model \cite{Hallaji2023}. In this setting, usually, the intruder disguises as a member of the federation and uses its artificially made data. As an example, reference \cite{HALLAJI2023110384} studies different variations of label poisoning in federated learning systems and mitigates these attacks by reformulating the problem as a label noise classification task.

Aside from the aforementioned mechanisms for data poisoning, the following attacks can be planned to take advantage of the blockchain backbone that most DFL methods rely on:

\textbf{Disrupting Communications:}
Blockchain is robust by nature against Denial of Service (DoS) attacks as there is no possible SPF that can be targeted. Nevertheless, Distributed DoS (DDoS) attacks can still slow down inter-node communications in the DFL system. DDoS agents continuously generate fake transactions and send them to the chain to fill up the blockchain buffer with spam data once the blockchain capacity is reached. This significantly postpones the inclusion of valid data into the upcoming blocks which puts the functionality of the blockchain under question. This is besides the permanent effect DDoS leaves on the DFL due to the data immutability of the blockchain.

\textbf{Degrading Detection Performance:} Using a targeted attack model, an intruder can inject certain triggers into the local model and obtain gradients in a way that the rest of the model remains intact \cite{DBLP:journals/corr/abs-1912-04977}. This backdoor attack results in sharing the poisoned parameters with the rest of the DFL participants and introducing the backdoor into their model as well. For instance, backdoors are injected into the DFL network in \cite{DBLP:journals/corr/abs-2202-02817} with two different mechanisms, namely label flipping \cite{pmlr-v119-rosenfeld20b} and pixel patch backdoor \cite{pmlr-v108-bagdasaryan20a}. The number of malicious nodes is restricted to prevent intruders from controlling the consensus (i.e., $2N_{corrupted}+2<N$, where $N$ and $N_{corrupted}$ indicate the number of all nodes and corrupted ones, respectively). A pixel pattern backdoor is also simulated in \cite{10.1145/3422337.3447837} for DFL with smart contract. The untargeted scheme of this attack which is also known as the Crashing DoS attack, follows the same process with the exception of degrading the performance of the entire model rather than for a specific class. While the untargeted version is easier to implement, it is also easier to detect. Note that the literature usually defines DoS differently for blockchain and FL. DoS in blockchain renders the global model inaccessible, whereas in FL, it makes the global model unusable. Reference \cite{9292450} presents an example of a data poisoning attack in DFL which is simulated with the aim of corrupting the global model performance. In this scenario, 30 percent of nodes are considered malicious, with the assumption that the number of malicious nodes does not change in time. Another case is tested in \cite{8622598} by generating forged gradients drawn from a Gaussian distribution by several Byzantine data holders (i.e., trusted nodes went rouge in the DFL network). Reference \cite{8994206} simulates this attack by filliping the class labels randomly on the training data. In contrast to poisoning attacks that primarily corrupt the training process, evasion or exploratory attacks occur during the inference phase after the model has been trained. Their goal is typically not to change the trained model, but rather to generate incorrect predictions or gather information about the model's characteristics.

\textbf{Causing Legal Problems:} By law, the inclusion of personal data is prohibited in some regions of the world. European general data protection regulation can be mentioned as a good example of this legal constraint. A malicious user, however, can poison the blockchain by inserting personal data into a data block and adding it to the chain which makes the DFL system non-compliant with the existing legal bounds. We call these adversaries privacy poisoning attacks. Despite its importance, not many research studies considering this type of attack in a DFL paradigm have been reported.

\begin{table*}[t]
\centering
\setlength{\tabcolsep}{17pt}
\caption{Characteristics and description of attacks in DFL. The ease of implementation, effectiveness, and defense feasibility for each attack is evaluated.}
\begin{tabular}{lcccc}
     \toprule
     Attacks & Target & Implementation & Effectiveness & Defensibility \\
     \midrule
     Backdoor & Performance & Challenging & High & Challenging \\
     DoS/DDoS & Performance & Simple & Moderate & Simple \\
     Gradient manipulation & Performance & Challenging & High & Challenging \\
     Training objective manipulation & Performance & Challenging & High & Challenging \\
     Evasion attacks & Performance & Simple & Moderate & Moderate \\
     Model inversion & Privacy & Simple & Low & Moderate \\
     Feature inference & Privacy & Simple & Moderate & Moderate \\
     Membership inference & Privacy & Challenging & High & Challenging \\
     Hijacking private key & Blockchain & Challenging & High & Challenging \\
     51\% attack & Blockchain & Challenging & High & Challenging \\
     Sybil attacks & Blockchain & Simple & High & Challenging \\
     Double spending & Blockchain & Challenging & High & Challenging \\
     Routing attacks & Blockchain & Simple & Moderate & Challenging \\
     Privacy poisoning & Blockchain/Privacy & Moderate & Moderate & Moderate \\
     Attacks on SC & Blockchain & Challenging & High & Moderate \\
     \bottomrule
\end{tabular}
\label{tab:attacks}
\end{table*}

\subsubsection{Model Poisoning}
Model poisoning is the process of maliciously controlling global model training. While data poisoning can be also used as a tool to cause model poisoning \cite{DBLP:journals/corr/abs-2202-02817}, model poisoning without data manipulation is also possible. As an example, one can simply change the objective of the local model to obtain poisoned gradients using valid data on the same model structure as the rest of the network \cite{DBLP:journals/corr/abs-1912-04977}. For instance, an additional term can be added to the objective to penalize sensitivity to malicious data \cite{pmlr-v97-bhagoji19a}. The same approach can be followed to deteriorate the overall performance in an untargeted manner (e.g., gradient manipulation). This case is also studied in \cite{9210138} where a set of Byzantine nodes are considered to generate and inject deviating gradients into the blockchain in order to degrade the shared DFL model via an outsider attack. A similar approach is used in \cite{9293091} to corrupt gradients using pointwise Gaussian noise.

\subsubsection{Routing Attacks}
Reliable network infrastructure is of paramount importance for DFL due to its integration with blockchain \cite{LI2020841, KHAN2018395}. This is while some of the utilized network protocols used by current internet service providers, such as the border gateway protocol, come with security flaws. Blockchain has no control over the network layer as it mostly works in the application layer. Thus, if the service provider network is breached, the routing of packets can be tampered with to either discard the transferring data blocks or change the blockchain structure. An example of a routing attack is a hack that took place in 2014 in which the intruder used the hijacked blocks to provide PoW and steal the associated rewards.

\subsubsection{Consensus Attacks}
The objective of these attacks is to obtain the majority of approvals in order to control the consensus mechanism in the DFL system. Depending on the number of agents or nodes involved in these attacks, they will be called 51\% or Sybil attacks. Sybil attacks indicate the scenario in which an intruder creates multiple fake nodes and claims to be more than one entity to impose a greater influence on the consensus. If intruders manage to take control of 51 percent of the nodes, the consensus mechanism is fully controllable. At this scale, the attack is known as 51\% attack in the literature \cite{9293091}. It is worth mentioning that consensus attacks are most effective when the blockchain associated with the DFL is at its early stage when the number of participants is still limited. PoW and PoS can drastically complicate consensus attacks for hackers.

\subsection{Attacks on Privacy}
Despite the decentralization of data in DFL, gradient and parameter information can be visible to system clients. Previously, we studied the effect of such information leakage on the performance of the DFL model. Here, we elaborate on the privacy risks associated with DFL networks.

\subsubsection{Model Inversion Attacks}
Model inversion can be used to approximate and reconstruct the private data of clients merely based on the classification model \cite{DBLP:journals/corr/abs-1912-04977}. An example of this attack is reported in \cite{DBLP:journals/corr/abs-2202-02817} where gradient leakage \cite{NEURIPS2019_60a6c400} of a global DFL model is used for reconstructing the user's local data. This process is often carried out using a DL-based generator. For instance, Generative Adversarial Networks (GAN) can be employed to perform model inversion by using the target model as the discriminator and training a generator network to minimize a cost that eventually converges and produces data samples similar to that of the targeted client. 

\subsubsection{Membership Inference Attacks}
Membership inference attacks are focused on revealing sensitive information associated with certain samples. Membership inference can disclose the membership of a sample to a certain class, or elicit attribute information \cite{8835269, 10.1145/3133956.3134012}.  An example of this attack can be found in \cite{9292450}, where an inference attack by observing the clients' updates in a DFL network and using the leaked gradients to infer record-level information of local datasets.

\subsubsection{Hijacking Private Key}
Securing private keys and public key using cryptography is one of the most critical aspects of blockchain which is the backbone of DFL. If there is any imperfection in the key signing mechanism, hackers can hijack a client's private key using their public key. Taking over a client's private key enables the hacker to gain full access to the corresponding data in the blockchain.

\subsubsection{Vulnerability of Smart Contracts}
Smart contracts, which are coded agreements utilizing blockchain technology for record-keeping, can be susceptible to security flaws. While they eliminate the need for intermediaries and provide immutable contracts, there is a risk associated with poorly coded smart contracts. These coding vulnerabilities create opportunities for attackers to identify flaws in the code and exploit them. By exploiting these weaknesses, attackers can potentially manipulate or extract unauthorized access to the contract's contents or associated assets. It is crucial to ensure thorough code review and rigorous testing to identify and address these security flaws to safeguard the integrity and trustworthiness of smart contracts.

\subsection{Analyzing the Viability of Attacks}
Table \ref{tab:attacks} provides an overview of the analyzed attacks, highlighting their key characteristics. When evaluating these attacks, it is crucial to assess their feasibility and effectiveness to design effective defense mechanisms and allocate appropriate resources to secure DFL systems. 

\subsubsection{Ease of implementation}
Implementing a backdoor attack is considered to be difficult because it requires extensive knowledge of the target DFL system, access to training data, and the ability to modify model parameters \cite{DBLP:journals/corr/abs-2007-05084}. Backdoor attacks often involve sophisticated techniques and may require compromising multiple clients or insiders with privileged access \cite{pmlr-v108-bagdasaryan20a, Xie2020DBA:}. Conversely, carrying out a DoS or DDoS attack is relatively less complicated. These attacks overwhelm the system with a high volume of requests or  malicious traffic, and readily available tools and techniques can be used to execute them \cite{10018276}.

Gradient manipulation necessitates a deep understanding of the DFL system's architecture, algorithms, and access to the communication channels. Tampering with gradients exchanged between clients and the server is a challenging task that requires careful execution to avoid detection \cite{9912299}. Similarly, manipulating the training objective, such as modifying the loss function or optimization process, is also categorized as hard. It demands an in-depth understanding of the system's algorithms, access to the training process, and the ability to modify the objective without disrupting the overall learning process.

On the other hand, attacks such as evasion attacks and model inversion are easier to implement. These attacks can be implemented using existing knowledge and techniques \cite{10.1145/2810103.2813677}. Other attacks, including feature inference and Sybil attacks, are also categorized as easy to implement. In contrast, attacks such as membership inference, hijacking private keys, 51\% attacks, consensus attacks, and double spending are challenging attacks \cite{CHEN2022100048,li_blockchain_2021}. These attacks require advanced knowledge, sophisticated attacks, and access to specific components or mechanisms of the DFL system. Implementing them successfully is a complex and resource-intensive task. Furthermore, privacy poisoning attacks are not overly complex or technically challenging, leading to a medium level of ease of implementation. The process may involve tampering with the data blocks and ensuring the inclusion of personal information, but it does not require advanced technical skills or extensive resources. Moreover, finding security holes in SC is a complex task as these protocols are usually well-tested before release, and sophisticated technical skills are required to hack SC under this condition \cite{CHU2023107221}.

\subsubsection{Effectiveness of Attacks}
In terms of attack effectiveness, the backdoor attack is classified as high. This attack has the potential to significantly impact the model's predictions, compromising the integrity and reliability of the DFL system \cite{pmlr-v108-bagdasaryan20a, DBLP:journals/corr/abs-2007-05084, Xie2020DBA:}. On the other hand, DoS/DDoS attacks are considered to have a moderate level of effectiveness. While they can disrupt the normal functioning of the system and cause temporary delays, their impact on the actual model parameters or the accuracy of the aggregated model may vary \cite{10018276}.

Gradient manipulation and training objective manipulation attacks are both considered to have a moderate level of effectiveness. Tampering with gradients or modifying the training objective can potentially influence the learning process and impact the accuracy of the aggregated model. However, their success depends on the robustness of the DFL system and the defense mechanisms in place \cite{9912299}.

Evasion attacks and model inversion attacks are also considered to have a moderate level of effectiveness \cite{10.1145/2810103.2813677}. Their impact on the overall model performance may be limited depending on the effectiveness of the defense mechanisms. Membership inference attacks, on the other hand, are classified as having a high level of effectiveness. Successfully inferring membership information about specific individuals participating in the DFL system can have severe privacy implications.

Attacks targeting the blockchain aspect of DFL, such as hijacking private keys, 51\% attacks, consensus attacks, Sybil attacks, double spending, and routing attacks, are generally considered to have a high level of effectiveness \cite{CHEN2022100048}. These attacks have the potential to compromise the security, integrity, and trust of the DFL system, depending on the specific vulnerabilities and defense mechanisms in place. Moreover, privacy poisoning attacks can have a moderate impact on compromising the privacy of participants in the DFL system. While the effectiveness of privacy poisoning attacks is not as high as some other attacks, it still poses a significant risk to the privacy of participants and the overall integrity of the system. In addition, exploiting the vulnerabilities of SC can lead to losing digital trust and massive financial loss. For instance, reports indicate that crypto investors nearly lost four billion dollars to hackers in 2022.

\subsubsection{Defensibility}
Backdoor attacks involve inserting malicious behavior into the training data or model, making it challenging to detect and defend against. Detecting and mitigating backdoor attacks often require advanced techniques and rigorous model verification \cite{8835365, li2021neural}.

DoS/DDoS attacks are classified as moderately defensible. While it may be challenging to completely prevent DoS/DDoS attacks, there are various mitigation strategies available, including anomaly detection. Such methods can reduce the impact of DoS/DDoS attacks and keep the system operational \cite{DORIGUZZICORIN2024103597, NIPS2017_f4b9ec30, 10.1145/2991079.2991125}.

Defending against gradient manipulation and training objective manipulation attacks is also categorized as moderately defensible. Implementing techniques like DP, robust aggregation algorithms, and secure communication protocols can enhance the system's resilience against these attacks \cite{9912299, 10.1145/1807167.1807247}. However, the effectiveness of the defense mechanisms may vary based on the sophistication of the attack and the quality of the defense techniques.

Evasion attacks and model inversion attacks are considered moderately defensible. Privacy-preserving techniques, such as DP or SMC, can help protect against these attacks \cite{10.1145/1807167.1807247, 10.1007/978-3-642-24178-9_9}. However, achieving strong defenses requires careful design and implementation, considering factors like attack vectors and data sensitivity.

Defending against membership inference attacks is classified as challenging due to the difficulty of protecting individual privacy in the FL setting. Advanced privacy protection techniques, such as privacy amplification and enhanced model aggregation protocols, are needed to effectively mitigate the risk of membership inference attacks \cite{NEURIPS2018_3b5020bb, 8555134}.

Regarding attacks on the blockchain aspect of DFL, the defensibility levels vary depending on the specific attack \cite{CHEN2022100048}. Hijacking private keys, 51\% attacks, and double spending attacks are challenging to defend against due to the vulnerabilities they exploit in the blockchain infrastructure. Implementing robust consensus mechanisms, multi-factor authentication, and encryption can enhance the defensibility against these attacks. Defending against consensus attacks, Sybil attacks, and routing attacks requires strong identity management systems, network monitoring, and reputation-based mechanisms. Defending against privacy poisoning attacks requires a combination of technical, legal, and regulatory measures, making it of medium difficulty in terms of defensibility. In addition, ensuring the security and integrity of SC has a moderate level of difficulty since it requires sophisticated code review and rigorous testing to identify and eliminate potential security holes and flaws.

\begin{table*}[t]
\centering
\caption{Characteristics and applicability of defense mechanisms. The final column denotes the specific types of attacks for which each defense mechanism is suitable.}
\label{tab:defs}
\begin{tabular}{llll}
\toprule
Defense mechanism & Description & Weakness & Used against \\
\toprule 
Homomorphic Encryption & Encrypts parameters & Computational overhead & Attacks on privacy\\
\midrule
SMC & Protects data with multiparty computation & Communication overhead & Attacks on privacy\\
\midrule
Differential Privacy & Perturbs parameters with noise & Accuracy loss & Attacks on privacy\\
\midrule
Anomaly Detection & Monitors updates to detect abnormalities & Backdoors can bypass it & Untargeted poisoning\\
\midrule
Robust Aggregation & Safeguards global aggregation against adversaries & Limiting assumptions & Poisoning attacks\\
\midrule
\multirow{2}{*}{Pruning} & \multirow{2}{*}{Randomly drops neurons from global model} & Computational overhead & \multirow{2}{*}{Backdoor attacks}\\
& & Performance loss \\
\midrule
TEE & A secure ecosystem for maintaining digital trust & Limited memory size & White box attacks\\
\midrule
\multirow{3}{*}{Zero-Knowledge Proofs} & \multirow{3}{*}{Uses unconnected bits of information to hide data} & Computational overhead & \multirow{2}{*}{Privacy attacks}\\
& & Communication overhead & \multirow{2}{*}{Poisoning attacks}\\
& & Complex implementation &  \\
\midrule
\multirow{2}{*}{Knowledge Distillation} & \multirow{2}{*}{Performs TL to indirectly train a smaller model} & Computational overhead & Privacy attacks\\
& & Not a stand-alone defense & Poisoning attacks\\
\midrule
\multirow{2}{*}{Regularization} & \multirow{2}{*}{Prevents global model from overfitting} & Sensitivity to hyperparameters& Privacy attacks\\
& & Computational overhead &  Poisoning attacks\\
\midrule
\multirow{2}{*}{Blockchain} & \multirow{2}{*}{Secures DFL communications} & Computational overhead & \multirow{2}{*}{Poisoning attacks}\\
& & Blockchain security issues & \\
\bottomrule
\end{tabular}
\end{table*}

\begin{table*}
	\centering
	\setlength{\tabcolsep}{8pt}
	\caption{Security analysis of state-of-the-art DFL methods. All models are robust against SPF. \Circle, \LEFTcircle, and \CIRCLE~denote not robust, partially robust, and robust, respectively. \checkmark and $\times$ indicate whether the specified technology is used or not. Attacks or technologies that are not included in this table are not studied in the listed methods.}
	\begin{tabular}{lccccccccc}
		\toprule
		Methods       & References & Blockchain & Encryption & Clients & Data poisoning & Model poisoning & Inference      \\ 
		\toprule
		BindaaS  &\cite{8943171}      &   \checkmark  & $\times$ & Honest & \Circle & \Circle & \Circle \\
		\midrule
		PIRATE &\cite{9210138}        &     \checkmark  & $\times$  & Semi-honest&  \CIRCLE & \Circle & \LEFTcircle \\
		\midrule
		BAFFLE  &\cite{9284684}       &      \checkmark   & $\times$   & Dishonest &\CIRCLE & \Circle & \CIRCLE \\
		\midrule
		BFLC  &\cite{9293091}         &         \checkmark    & $\times$   & Semi-honest & \CIRCLE & \Circle  &  \Circle     \\
		\midrule
		LearningChain &\cite{8622598} & \checkmark  & DP & Dishonest & \LEFTcircle & \LEFTcircle & \CIRCLE\\
		\midrule
		Biscotti &\cite{9292450}      &   \checkmark & DP   & Dishonest & \CIRCLE & \LEFTcircle & \CIRCLE\\
		\midrule
		Blade-FL & \cite{9664296}     &     \checkmark &  DP & Honest & \Circle &  \Circle& \LEFTcircle \\
		\midrule
		BEAS &\cite{DBLP:journals/corr/abs-2202-02817}         &     \checkmark   & DP    & Dishonest & \CIRCLE & \CIRCLE    & \CIRCLE     \\ 
		\midrule
		Swarm Learning & \cite{warnat-herresthal_swarm_2021} &\checkmark  & $\times$ & Dishonest & \Circle & \Circle&  \Circle\\
		\midrule
		SL+HE & \cite{10.1007/978-3-030-95391-1_32} &\checkmark  & HE & Dishonest & \Circle & \Circle&  \CIRCLE\\
		\midrule
		BrainTorrent & \cite{roy2019braintorrent}  & $\times$ & $\times$ & Honest & \Circle & \Circle  & \Circle\\
		\midrule
		P2P-FL &\cite{DBLP:journals/corr/abs-1901-11173} & $\times$& $\times$ & Honest &  \Circle & \Circle  & \Circle\\
		\midrule
		ReputationDFL  & \cite{8994206} & \checkmark & $\times$ &  Dishonest & \LEFTcircle & \LEFTcircle & \Circle\\
		\midrule
		BlockFLA & \cite{10.1145/3422337.3447837} & \checkmark & $\times $ & Dishonest & \LEFTcircle & \LEFTcircle & \Circle \\
		\midrule
		DFedForest & \cite{9284805} & \checkmark & $\times$ & Dishonest & \LEFTcircle & \LEFTcircle & \Circle \\
		\midrule
		Bift & \cite{9650783} & \checkmark & $\times$ & Dishonest & \LEFTcircle & \LEFTcircle & \Circle\\
		\midrule
		Verifiable DFL & \cite{FANG20221} & \checkmark & $\times$& Semi-honest & \Circle & \Circle & \CIRCLE\\
		\midrule
		DFL for Healthcare & \cite{10.1145/3426474} & \checkmark & DP, SMC & Semi-honest & \Circle & \Circle & \CIRCLE \\
		\midrule
		FLChain & \cite{8905038} & \checkmark & $\times$& Honest & \Circle & \Circle & \Circle  \\
		\midrule
		VFChain & \cite{9321132} & \checkmark & $\times$ & Semi-honest & \Circle & \Circle & \LEFTcircle \\
		\bottomrule
	\end{tabular}
	\label{tab:analysis}
\end{table*}

\section{Defense Mechanisms}
Various defense mechanisms are proposed to fortify DFL against privacy and performance-related threats. While this section reviews the literature on DFL defense mechanisms, their characteristics are summarized in Table \ref{tab:defs}. In addition, Table \ref{tab:analysis} specifies mechanisms that each reviewed DFL method employs. 

\subsection{Privacy Preserving}
Despite the wide diversity of previous efforts on safeguarding FL and blockchain privacy, suggested methods typically fall into one of these three categories: 1) Homomorphic Encryption (HE), 2) Secure Multiparty Computation (SMC), and 3) DP. The following discussion goes through each of these groups.

\subsubsection{Homomorphic Encryption}
By processing on cyphertext, HE is commonly used to secure the learning process. Clients can use HE to perform arithmetic operations on encrypted data (i.e., ciphertext) without having to decode it. HE has three major variations that differ in arithmetic complexity and flexibility, namely full, partial, and substantial HE \cite{OGBURN2013502}.

Fully HE is capable of doing arbitrary calculations on the encrypted data \cite{10.1145/1536414.1536440}. This is while partially HE can only execute one operation (e.g., addition or multiplication), and substantially HE can do several operations \cite{10.1007/978-3-642-32009-5_38, 10.1145/359340.359342, 10.1007/3-540-48910-X_16}. The latter, on the other hand, has a restricted amount of additions and multiplications. While full HE offers greater flexibility, it is inefficient when compared to other forms of HE \cite{10.1145/1536414.1536440}. As an example, \cite{10.1007/978-3-030-95391-1_32} shows that swarm learning updates are encrypted using partial HE to defend DFL against inference attacks.

Despite the benefits of HE, executing arithmetic on the encrypted integers increases the memory and processing time costs. Moreover, non-linear estimations in statistical models demand approximating polynomials which make it important to find a balance between utility and privacy \cite{10.1145/2857705.2857731, info:doi/10.2196/medinform.8805}. In \cite{8241854}, for instance, additively HE is used to secure distributed learning by securing model changes and maintaining gradient privacy. Another example is \cite{DBLP:journals/corr/abs-1711-10677}, which uses an additively homomorphic architecture to defeat honest-but-curious adversaries using federated logistic regression on the encrypted vertical FL data. However, the overburdening of the system with additional computational costs is a typical downside of such systems.

\subsubsection{Secure Multiparty Computation}
Secure Multiparty Computation (SMC) \cite{4568388}, is a distributed cryptography technique in which several entities participate in the estimation of a function. The distinct feature of SMC is securing each participant's data by creating a set of random values that are not equal to the participant's data, and sending them to the rest of the parties for local calculation of the function. The outputs of these functions then will be averaged to obtain the desired estimation. In this scheme, the data on the participant's side is meaningless, and the function estimations are only usable once they are averaged. For instance, SMC was used in \cite{7958569} for private model training. It is worth mentioning that SMC is followed by excessive communication and computational cost. It has been also mentioned that SMC is best to be coupled with DP to secure the communications \cite{10.1145/1807167.1807247,10.1007/978-3-642-24178-9_9}.

In conclusion, SMC usage in large-scale DFL could be inefficient due to the dramatic rise in communication and processing costs. Secondly, encryption-based solutions corresponding to these functions must be properly defined and implemented \cite{10.1145/3243734.3243760, DBLP:journals/corr/abs-1901-00329}. Finally, all cryptography-based protocols preclude an audition phase of the received updates by the shared model, hence, leaving holes for rogue users to exploit.

\subsubsection{Differential Privacy}
The idea of DP is to inject random noise into the generating updates so that the data interpretation becomes infeasible for malicious entities. DP is primarily used to safeguard DFL communications against privacy attacks (e.g., inference attacks); however, the literature also shows that DP can be also beneficial against data poisoning attacks as these attacks are usually designed based on the communicated gradients \cite{10.5555/3367471.3367701, 10.1145/2976749.2978318, DBLP:journals/corr/abs-1712-07557}. Biscotti \cite{9292450} couples DP with a secure aggregation technique \cite{10.1145/359168.359176} to safeguard DFL systems against poisoning and inference attacks. Another report shows the effectiveness of pruning-based DP in safeguarding horizontal DFL systems \cite{DBLP:journals/corr/abs-2202-02817}. DP in \cite{8622598} is implemented through perturbing local gradients using the exponential mechanism \cite{10.1007/978-3-540-79228-4_1} and a predefined probability density function before uploading them into the blockchain and broadcasting them across the DFL network. Similarly, Blade-FL \cite{9664296} adds noise to the gradients prior to the encapsulation phase, albeit using a Gaussian distribution.

In contrast to HE and SMC whose main disadvantage was communication overhead, DP has a lower computational cost and does not overburden the system in this sense. Instead, DP comes at the cost of deteriorating the model quality. This is mainly because the injected noise can potentially add up to the noise within the constructed model. Another issue of concern in conventional DP is the cumulative privacy loss resulting from iterative training processes that utilize local data from multiple individuals or sources. These iterations are crucial for enhancing the accuracy and performance of trained models. However, with each iteration, a certain degree of privacy loss is introduced, and this loss accumulates over time, potentially reaching a significant level. Researchers have dedicated efforts to address this problem by exploring various approaches, including subsampling \cite{NEURIPS2018_3b5020bb} and privacy amplification by iteration \cite{8555134}. These techniques aim to mitigate cumulative privacy loss and improve the overall privacy guarantees of DP schemes. Since these methods derive a tight upper bound of cumulative privacy loss, they can also be applied to preserve model utility even in cases where the gradients are perturbed by noise. Moreover, DP provides resistance to poisoning attempts due to its group privacy trait. As a result, as the number of attackers increases, this defense will reduce significantly. As mentioned before, all privacy-preserving methods have a set of advantages and disadvantages, and thus, there is no perfect alternative for DP. In other words, DP, HE, and SMC each result in a different privacy-utility trade-off. As a solution, hybrid approaches can be implemented to build privacy protocols that are more robust than using only one of these methods \cite{10.1145/3338501.3357371, 10.1145/3338501.3357370}. As an example, SMC is combined with DP to balance the trade-off between them \cite{10.1145/3338501.3357370}. This combination offsets excessive noise injection when the number of clients is growing while preserving the desired rate of trust.

Gaussian DP is a specific case within the functional DP ($f$-DP) framework, which characterizes privacy through hypothesis testing \cite{10.1111/rssb.12454}. In $f$-DP, a randomized algorithm is considered to satisfy privacy if the difficulty of distinguishing between two neighboring datasets (quantified by a trade-off function) is element-wise larger than a convex and non-increasing function $f$. When this function f is constructed using two Gaussian distributions, the resulting form of $f$-DP is termed Gaussian Differential Privacy. In other words, Gaussian DP defines privacy guarantees by examining the distinguishability of neighboring datasets through a hypothesis testing approach, with the specific choice of a trade-off function derived from Gaussian distributions. This provides a mathematical and analytical framework for evaluating privacy in the context of private data analysis. For instance, \cite{pmlr-v130-zheng21a} applies Gaussian DP in an FL system to provide record-level privacy guarantees for each client. Building upon the ability of this approach to handle composition and subsampling, \cite{Bu2020Deep} extends the work to apply DP to SGD and Adam \cite{Kingma2015} optimizers.

In the context of applying DP to FL, two privacy notations, namely user level and instance level privacy, can be found in the literature. User-level privacy in DP ensures that clients' data remains private and the privacy of the global model remains intact, even if an adversary removes a client or its data from the aggregation process \cite{brendan2018learning}. Additionally, adversaries are unable to determine whether a client has participated in the training. On the other hand, employing DP at the record level offers privacy guarantees for each individual record within a client's data, protecting them against potential adversaries \cite{pmlr-v130-zheng21a}. Instance-level DP is often perceived as offering a relatively weaker privacy guarantee, as adversaries may potentially determine whether a user participated in the training or not. However, there are cases where discerning user participation is not crucial, and the primary concern is maintaining data privacy. Considering the trade-off between privacy and utility in differential privacy, it is prudent to adopt instance-level privacy in such scenarios to achieve improved performance while avoiding overly stringent privacy constraints.

DP can be centralized, local, or distributed. In centralized DP, the noise addition is performed via a server, which makes it impractical in DFL. On the other hand, local \cite{6736718} and distributed DP \cite{10.1145/2873069, dwork2006our} both assume that the aggregator is not trusted which perfectly complies with the DFL paradigm. In the local variant, participants inject noise in their estimated gradients before sharing them over the blockchain. However, research on local DP indicates its inability to provide a privacy guarantee on large-scale and heterogeneous models with numerous parameters \cite{DBLP:journals/corr/abs-2009-05537,papernot2017semisupervised}. In DFL, the injected noise should be calibrated to ensure successful DP. Despite the appealing security qualities of local DP, its practicality becomes questionable when dealing with an immense number of users. 

Distributed DP combines cryptographic techniques to provide the benefits of both local and centralized DP without compromising the clients' privacy \cite{10.5555/3327757.3327856, 10.1145/2873069}. As a result, it avoids putting faith in any server and is more effective in that sense. Decentralized DP, in theory, has the same benefit as the centralized variant because the overall quantity of noise is similar for them. The concept of distributed DP alludes to the notion that the required amount of noise is derived from several individuals \cite{dwork2006our}. 

\subsection{Model Robustness}
Defenses are classified into two types: proactive and reactive. Proactive defense is a low-cost method of anticipating attacks and associated consequences. The reactive defense operates by detecting an invasion and taking preventative steps. In the production environment, reactive defense is often deployed as a patch-up. DFL presents multiple additional attack surfaces throughout training, resulting in complicated and unique countermeasures. In this part, we will look at some of the most common types of DFL defensive tactics and investigate their usefulness and limits.

\subsubsection{Anomaly Detection}
Anomaly detection methods have a long history in real-time identification of data attacks in intelligent systems \cite{9928313}. In centralized FL, anomaly detection is often performed on the server to identify malicious updates that can lead to model or data poisoning \cite{NIPS2017_f4b9ec30, 10.1145/2991079.2991125}. It is worthwhile to mention that this approach usually works best against untargeted attacks. In the DFL paradigm, however, anomalies can take place either due to fraudulent transactions or changes in blockchain networks such as network division and blockchain fork. Each of these issues is separately studied in the literature and there is no anomaly detection framework designed to handle both. For instance, \cite{MORISHIMA2021107087} proposes a fast-paced detection scheme for monitoring transactions using a subgraph-based approach. The designed detection method is optimized for parallel processing GPU acceleration. On the other hand, another work, \cite{9201454}, designs a distributed anomaly detection scheme, called BAD, to combat eclipse attacks in blockchain networks by re-connecting unauthorized forks to the P2P network. 

The majority of the anomaly detection methods for FL security are proposed with respect to the centralized architecture of the FL \cite{10.1145/2991079.2991125, 10.5555/1387709.1387716, 8418594, li2020learning}, and anomaly detection in DFL is only studied in a limited number of works. As an example, BEAS \cite{DBLP:journals/corr/abs-2202-02817} employs two anomaly detection protocols, namely Multi-KRUM \cite{NIPS2017_f4b9ec30} and FoolsGold \cite{8637481} to detect poisoning attacks and identifying Sybil groups (i.e., group of malicious nodes coordinating a cyber-attack), respectively. The former monitors change in variance or performance of generated gradients with respect to the majority of updates. The latter, on the other hand, detects Sybil groups based on the correlation between the generated updates, as it is expected for malicious nodes to generate highly similar gradients. Anomaly detection in high-dimensional space poses challenges, leading some detectors to suggest using dimensionality reduction to make monitoring easier. However, reducing the dimensionality size may result in information loss \cite{9609642}, even with advanced techniques preserving data characteristics.

\subsubsection{Robust Aggregation}
The aggregation algorithm used in an FL system should tolerate communications disturbances, client dropout, and incorrect model updates on top of hostile participants \cite{9026922}. Extensive research has been dedicated to advancing robust aggregation in centralized FL \cite{pillutla2019robust, DBLP:journals/corr/abs-2009-08294}. Similarly, in DFL networks, the security of the aggregation phase is of paramount importance. Under a DFL paradigm, the aggregation phase is mainly secured by employing the same techniques used in FL in combination with the consensus mechanism of the utilized blockchain scheme \cite{9650783}. These protocols (e.g., PoW, PoS, SC) control which nodes can participate in the model aggregation phase for both generating updates and aggregating parameters. On the downside, many of the advanced Byzantine-robust aggregations rely on assumptions that are either unrealistic or incompatible within the context of FL \cite{DBLP:journals/corr/abs-1912-04977}.

\subsubsection{Pruning}
Even though pruning is not specific to DFL and can be applied to any neural network structure, it can help in eliminating backdoors. The idea of pruning is to drop some neurons in order to enhance the efficiency and accuracy of the network. However, this feature makes the parameter usage somewhat unpredictable for attackers which complicates injecting backdoors into a neural network since inactive neurons will be removed eventually in the network. While pruning has been mainly studied for FL \cite{DBLP:journals/corr/abs-1812-07210, DBLP:journals/corr/abs-1909-12326}, current literature on pruning under a DFL paradigm is very limited \cite{DBLP:journals/corr/abs-2202-02817}. In particular, BEAS \cite{DBLP:journals/corr/abs-2202-02817} makes use of gradient pruning \cite{NEURIPS2019_60a6c400} in DFL to facilitate DP and complicating model poisoning for attackers in the system. Nevertheless, since these techniques are mainly used at the edge of the network, most pruning approaches should be adaptable to DFL as well. 

Pruning comes with a set of drawbacks. One drawback is the potential loss of model capacity, resulting from the removal of connections or components. Another concern is the sensitivity of pruning to initialization and training, necessitating careful fine-tuning and experimentation for optimal outcomes. Moreover, the pruned model may struggle to generalize effectively to new data, limiting its real-world applicability. Lastly, the computational overhead increases due to the additional resources and time required during the pruning and retraining stages. Considering these factors is vital when assessing the viability of pruning as a defense mechanism against backdoors.

\subsubsection{Trusted Execution Environment}
The blockchain backbone of DFL requires the data to be replicated on each node which makes it challenging to keep smart contracts confidential. TEE (Trusted Execution Environment), on the other hand, is a tamper-resistant ecosystem that can be used to maintain digital trust between nodes of a distributed network \cite{9411833}. TEE is often referred to as a secure and isolated part of the processor that requires all codes and data signatures to be verified with respect to the designer's expectations.  The validity of a participating device in a TEE Authentication should be checked by the connected service with which it is attempting to enroll. Until the matching party provides a message, the status of code execution stays hidden. The execution route of the code cannot be changed until it takes explicit input or a validated interruption. Data stored on and processed by participants is safe, and interactions between various parties are carried out in a secure manner. The TEE is in charge of all data access privileges. Cryptographic technologies are used to secure TEE communications. Only the TEE secure environment stores, maintains, and uses private and public encryption keys. The TEE can show a remote client what code is presently being executed as well as the starting state. TEE can aid in resolving a key challenge for FL security since it is becoming progressively important in securing the central server and clients against hackers and preventing data theft.

TEE is coupled with FL to safeguard against algorithmic attacks \cite{CHEN202069, 10.1145/3458864.3466628}. TEE can hide model parameters on local devices so that the model is not accessible to the attacker. Under this condition, the attacker is only able to launch black box attacks on the system. Nevertheless, TEE often suffers from a limited memory size, that is only a limited part of the model can be secured with this approach \cite{10.1145/3458864.3466628}. It has been also suggested that TEE can secure smart contract data in a blockchain which is the main vulnerability of this architecture \cite{DBLP:journals/corr/abs-1805-08541}. Theoretically, the same concept can be applied to DFL. However, current literature lacks an experimental study on the integration of DFL and TEE.

\subsubsection{Zero-Knowledge Proofs}
The origin of zero-knowledge proofs goes back to the mid-1980s \cite{doi:10.1137/0218012}. This cryptographic approach allows a verification process that does not involve data exchange between parties \cite{6547113}.  This process often involves using unconnected bits of information to keep data private during the verification. 

As an example, zero-knowledge proofs have the potential to be utilized in DFL to verify the authenticity of the features used by clients for training and generating updates. While zero-knowledge proofs offer a promising avenue for enhancing secure update monitoring, additional research is necessary to identify challenges in constructing and implementing their modules. Notably, zero-knowledge proof protocols generally maintain their performance regardless of the data volume. 

Zero-knowledge proofs offer advantages for enhancing secure update monitoring in DFL. However, their adoption also entails certain drawbacks. These include computational overhead due to resource-intensive operations, the complexity of implementation requiring meticulous attention to cryptographic techniques, potential scalability issues in large-scale DFL systems due to communication complexity, and the trust assumption associated with the setup phase \cite{cryptoeprint:2021/730}.

\subsubsection{Knowledge Distillation}
Knowledge distillation is a fundamental algorithm in federated distillation \cite{li2019fedmd}. The goal of knowledge distillation is to perform TL from a large teacher model ($\mathcal{T}$) to a compact student model ($\mathcal{S}$) without sacrificing performance significantly. Smaller models have fewer parameters and are less susceptible to overfitting, making them more resilient against attacks. This increased resistance makes it more challenging for attackers to reverse-engineer or manipulate the model. Nonetheless, for certain threats such as backdoor attacks, knowledge distillation alone may not be directly effective, as these attacks typically involve modifying the training data or model parameters to embed a hidden trigger. However, knowledge distillation can be used as part of a broader defense strategy to mitigate the impact of backdoor attacks. For instance, reference \cite{li2021neural} mitigates backdoor attacks by utilizing knowledge distillation in deep neural networks. This approach independently fine-tunes $\mathcal{T}$ on a clean subset and uses it to clean backdoored $\mathcal{S}$. To do so, this method tries to align intermediate-layer attention in $\mathcal{S}$ with that of $\mathcal{T}$. Another example trains a $\mathcal{T}$ ensemble on disjoint training subsets and trains $\mathcal{S}$ based on aggregated noisy voting among $\mathcal{T}$ models \cite{papernot2017semisupervised}. The goal of this approach is to provide a privacy guarantee for training data. In DFL, this idea translates into sharing the knowledge of a model rather than the parameters which improves FL's robustness against both poisoning and privacy attacks. In addition to the mentioned security advantages, knowledge distillation also results in communication and computation efficiency in DFL. Exchanging model parameters becomes burdensome when communication resources are limited, especially for contemporary big deep neural networks. In this sense, federated distillation \cite{DBLP:journals/corr/abs-2009-05537} is an appealing FL option since it only transmits model outputs which are often considerably less in size than the model sizes. 

\subsubsection{Regularization}
In DFL, the computational model that is being distributively updated is most likely a DL structure. One approach to mislead this model is to make it overfit by introducing malicious samples which may in turn lead to membership inference. Regularization techniques such as $L_2$ regularizer  \cite{7958568} and dropout \cite{DBLP:journals/corr/abs-1806-01246} can eliminate the effect of such malicious samples to a great extent. Reference \cite{raghunathan2018certified} designs a semidefinite relaxation method that generates a differentiable certificate for network robustness and optimizes it alongside network parameters to encourage robustness against all attacks. Furthermore, determining suitable hyperparameters, such as regularization strength, for regularization techniques can be a difficult task, and incorrect choices may yield undesirable results. Moreover, the use of certain regularization methods with intricate terms or penalties can introduce considerable computational complexity, especially in scenarios involving extensive datasets or intricate models, resulting in prolonged training durations.

\section{Verifiable DFL}
Defense mechanisms reviewed in the previous section mainly combat an adversary after a malicious attack is launched by an intruder. Nevertheless, another approach to combat these threats is to prevent malicious parties to take part in the training process. A DFL system that can verify the trustworthiness of its clients is called verifiable.

A DFL system is considered verifiable if clients of a network can prove to each other that the given task has been carried out without compromising privacy \cite{ijcai2022p792}.
Security, transparency, and automation of blockchain can enhance the verifiability of DFL \cite{10.1145/3426474} which in turn prevent malicious activities. In this structure, some nodes are in charge of generating updates and training the model (i.e., trainers) and groups of participants verify the generated updates (i.e., workers) with respect to consensus protocol. If the update is verified, the worker who completes this task first will create a new block. Then, if the majority of the workers approve the content of the created block, it will be appended to the chain. At this point, DFL clients can update their models using this new block of information. This process is repeated periodically to keep the DFL model up to date.

\subsection{Trustable DFL Trainers}
A vulnerability of DFL systems is that dishonest trainers can upload maliciously crafted updates to the blockchain. For instance, BlockFL \cite{8733825} verifies the submitted updates through a random selection of workers before letting the update take effect. In BlockFL, each client is associated with a random worker who is rewarded when evaluating the updates prior to the aggregation. The evaluation criteria are designed based on the relationship between the data size and the elapsed time for generating the update.

Another approach for verifying trainers is to use proof of correctness. Examples of this are presented in \cite{8905038, DBLP:journals/corr/abs-1906-10893}, where produced parameters are paired with their proof of correctness which is used by registered workers for determining the approved updates. This approach uses verifiable random functions \cite{DBLP:journals/corr/abs-1906-10893} or reliability ranking \cite{8905038} to select a worker whose aggregation results will be appended to the blockchain. The trainer's reliability, which is an indicator of its performance, will affect its chance to take part in the training process thereafter.

\subsection{Trustable DFL Workers}
Assuming that all trainers are trustworthy, the aggregated model could still be compromised if the responsible worker for the model aggregation is unreliable, that is, it does not follow the defined protocol. VFChain \cite{9321132} addresses this issue by using trainer signatures that are appended to the uploaded block by the trainer. To verify the aggregated model by the responsible worker, a committee is randomly selected among other workers to evaluate the aggregated model with respect to the verifying contract. The members of this committee are continuously updated in each training round. The blockchain preserves signatures and their corresponding aggregated models of each training round which makes all records tamper-resistant and transparent to all network members. It is worth mentioning that this structure is only compatible with semi-honest trainers.

\section{Future Research Directions}
Despite the advancements of DFL in the past five years, this paradigm is fairly new and still has a lot of potential to be improved. From a security standpoint, many of the available countermeasures in the literature are either studied for FL or blockchain, and their effectiveness in DFL remains unclear. Here, we outline future development requirements that we believe will be promising for DFL in this sense.

\subsection{Weakness of Privacy-Preserving Techniques}
As mentioned before, privacy-preserving techniques often trade excessive computational burden for enhancing privacy (e.g., HE, SMC). Furthermore, the implementation of cryptographic techniques such as SMC is sensitive and can imperil privacy if not defined properly. DP, on the other hand, can affect the accuracy of the aggregated model since the utilized noise can leak into the model. These issues are beside the fact that the audition phase in these techniques can lead to security holes. Hence, despite the history of privacy-preserving techniques, they all have imperfections that require further research and adaptation to the DFL paradigm.

Addressing these challenges requires ongoing research and adaptation of privacy-preserving techniques to suit the specific requirements of DFL. Developing optimized implementations of cryptographic methods, such as SMC, can help alleviate the computational overhead and address sensitivity concerns. Paying attention to secure parameter configurations and robust cryptographic protocols is essential to minimize privacy risks. Exploring advanced noise reduction techniques within DP can strike a balance between preserving privacy and maintaining the accuracy of the model \cite{pmlr-v130-zheng21a, DBLP:journals/corr/abs-2009-05537}. Robust auditing mechanisms need to be established to ensure the security of privacy-preserving techniques and safeguard against potential vulnerabilities \cite{8905038, 9321132}. Moreover, investigating hybrid approaches that integrate multiple privacy-preserving techniques can offer a compromise between privacy and computational efficiency \cite{10.1145/3426474}.

Continued exploration and refinement of privacy-preserving techniques in the DFL domain will enhance privacy and security while minimizing the impact on computational performance. This ongoing research and innovation will pave the way for more effective and privacy-aware DFL systems.

\subsection{Optimizing Defense Mechanism Deployment}
Devising defense mechanisms in DFL presents specific challenges that need to be addressed. One key challenge is the requirement for additional computational power, which may not always be feasible, particularly in resource-constrained environments like mobile networks \cite{10268062}. The limited computational resources available can pose limitations on the implementation of defense mechanisms. 

Another challenge is the diverse nature of threats in DFL. Different threats often require different countermeasures. Designing a comprehensive defense mechanism that addresses a wide range of threats can be complex and challenging. It requires analyzing various threat models, understanding their unique characteristics, and developing tailored defense strategies accordingly.

When considering DFL deployed on blockchain, resource and computational constraints specific to blockchain systems must also be taken into account \cite{9515771, 9635590}. The scalability of defense mechanisms becomes a critical aspect to consider. Strategies that optimize smart contract execution or leverage off-chain computations can help ensure efficient operation within the resource constraints of blockchain platforms.

Continuous evaluation, benchmarking, and improvement of existing defense mechanisms are vital. The field of defense mechanisms in DFL is dynamic, with emerging threats and advancements in technology. Regular evaluation and improvement are necessary to keep up with evolving challenges and maintain effective protection.

Overall, addressing the challenges of additional computational power, diverse threats, resource constraints in blockchain, and the optimal deployment of defense mechanisms are critical areas for further research and development in the field of DFL. By tackling these challenges and exploring the proposed solutions, researchers can advance the field and enhance the security and resilience of DFL systems.

\subsection{Blockchain-Related Security Issues}
The decentralization and traceability offered by DFL introduce unique challenges related to blockchain-related threats. These threats specifically exploit the structure of the blockchain, posing risks to the overall security of DFL systems. Consequently, there is a need for research and development aiming at creating more secure blockchain structures that can enhance the resilience of DFL ecosystems against these attacks.

One key challenge is the susceptibility of DFL to cyber-attacks that target the underlying blockchain infrastructure. Adversaries may attempt to manipulate or tamper with the blockchain, compromising the integrity and reliability of the system. This poses a significant risk to the confidentiality and privacy of participants' data in DFL. To address these challenges, research efforts should focus on the development of more secure blockchain structures tailored to the specific requirements of DFL. This involves exploring techniques such as improved consensus mechanisms, advanced cryptographic protocols, and enhanced smart contract design \cite{9347812}. These solutions aim to bolster the security of the blockchain layer, mitigating the vulnerabilities that can be exploited by malicious actors. Additionally, advancements in blockchain technology, such as the integration of privacy-preserving techniques and zero-knowledge proofs, can contribute to building a more attack-resilient DFL ecosystem. By leveraging techniques that enhance data privacy and confidentiality, DFL systems can better withstand attacks and protect sensitive information \cite{9292450, 10.1145/3426474, DBLP:journals/corr/abs-2202-02817}. Furthermore, the development of robust monitoring and auditing mechanisms is crucial for detecting and mitigating blockchain-related threats in DFL. Implementing effective monitoring systems can help identify suspicious activities and ensure the integrity of the blockchain \cite{8905038}. Auditing mechanisms can verify the correctness and security of the system, providing transparency and accountability in DFL deployments.

\subsection{Heterogeneity of Decentralized Federated Learning}
DFL convergence in a distributed ecosystem (e.g., blockchain) is not guaranteed when various clients have diverse computation power (i.e., heterogeneous users). Limited processing power and poor connectivity can delay the training and communication of some nodes, delaying the global aggregation. Disregarding this heterogeneity results deteriorates the overall efficiency of DFL training.

The heterogeneity of clients in DFL not only poses challenges to convergence and efficiency but also gives rise to security issues \cite{li2020federated}. In an asynchronous DFL setting, where clients operate with different computation power and connectivity, ensuring secure aggregation becomes challenging. Asynchronous DFL is not compatible with secure aggregation protocols because all clients need to be incorporated in the aggregation step when using secure aggregation. This also limits the ability to perform centralized or distributed DP. Consequently, the use of local DP techniques becomes the viable option in asynchronous DFL scenarios, potentially compromising the level of privacy protection and increasing the risk of information leakage during the aggregation process. These security concerns highlight the need for innovative approaches that strike a balance between achieving privacy-preserving aggregation and accommodating the heterogeneity of clients in DFL \cite{9005992}. By exploring novel cryptographic techniques, adaptive privacy mechanisms, and secure communication protocols, it is possible to address the security challenges that arise from the heterogeneity of DFL and ensure the confidentiality, integrity, and privacy of the FL process \cite{9066954, pmlr-v97-wu19c}.
	
\subsection{Challenges in Verifiable Decentralized Federated Learning}
Joining a DFL network requires a set of verifiers to evaluate the trustworthiness of the joining party. Since the joining party will not disclose their information for privacy reasons, the evaluation process often involves monitoring the performance of the joining party in the training process. Nevertheless, this evaluation scheme is not efficient as it exhausts computational and communication resources. Another challenge in this domain is to gain insight into the misbehavior of clients based on the history of their behavior. It is important to discover how the verification results can be employed to facilitate this task.

One possible solution is to employ privacy-preserving techniques during the trust evaluation process \cite{FANG20221, warnat-herresthal_swarm_2021}. Techniques such as SMC or HE can enable verifiers to evaluate the performance and trustworthiness of the joining party without directly accessing their sensitive information. This way, privacy is maintained while still obtaining the necessary insights for trust evaluation. Furthermore, optimizing the evaluation process can enhance its efficiency \cite{9321132}. Implementing techniques that reduce the computational and communication overhead, such as lightweight monitoring mechanisms or selective data aggregation, can help alleviate the resource exhaustion issues associated with the evaluation scheme. By carefully designing evaluation protocols and employing efficient algorithms, the overall efficiency of the process can be improved without compromising the evaluation's effectiveness.

In terms of addressing misbehavior based on historical behavior, leveraging machine learning techniques and anomaly detection algorithms can be beneficial. By analyzing the behavior patterns and performance metrics of clients over time, it is possible to identify potential anomalies or deviations that may indicate misbehavior. Establishing a comprehensive system for tracking and monitoring client behavior, coupled with intelligent analysis and detection mechanisms, can aid in detecting and addressing misbehavior in a timely manner. Moreover, incorporating reputation systems within the DFL network can provide additional insights into the trustworthiness of participating clients \cite{8832210, 8994206}. Verifiers can leverage the verification results and feedback from previous collaborations to build reputation scores for each client. These scores can serve as indicators of past behavior and can be used as a basis for decision-making during the trust evaluation process.
	
\section{Conclusion}
The integration of FL and blockchain technologies has alleviated the need for a server in the network. This is followed by a number of advantages such as efficient communication and the elimination of a single point of failure in the federation. While there are a limited number of surveys on the security analysis of FL, since the previous studies were all based on centralized architectures, further analysis is required to study the new paradigm of DFL from a security perspective. This work first reviewed common trends and preliminaries of FL and blockchain. It then performed a security analysis on DFL by identifying possible threats and defense mechanisms in such systems.

\ifCLASSOPTIONcaptionsoff
\newpage
\fi

\bibliographystyle{IEEEtran}
\bibliography{refs}

\begin{IEEEbiography}[{\includegraphics[width=1in,height=1.25in,clip,keepaspectratio]{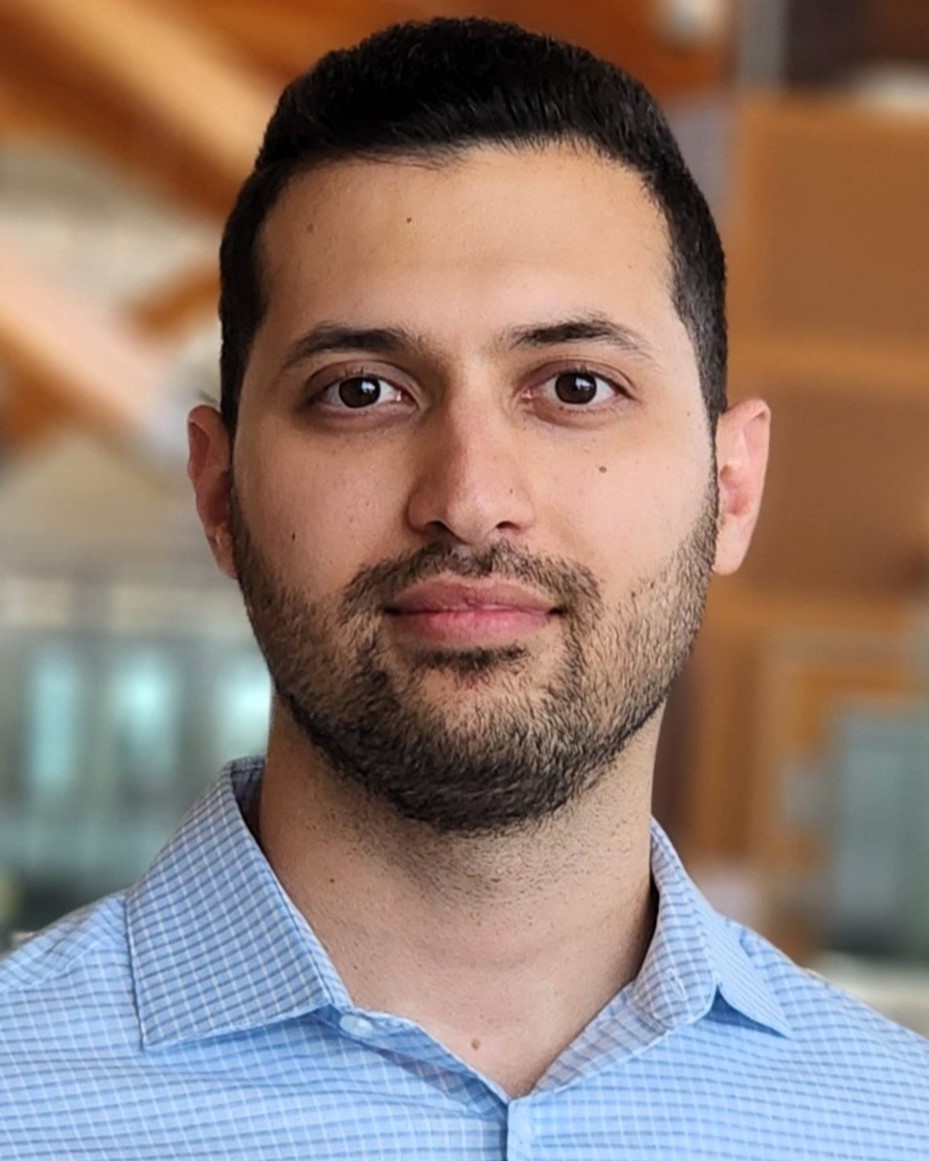}}]{Ehsan Hallaji}
(Graduate Student Member, IEEE) received the B.Sc. degree in software engineering from Shahid Rajaee University, Tehran, Iran, in 2015, and the M.A.Sc. degree in electrical and computer engineering from the University of Windsor, Windsor, ON, Canada, in 2018, where he is currently pursuing a Ph.D. degree with the Department of Electrical and Computer Engineering. His current research interests include machine learning, data mining, federated learning, and cybersecurity. He is a reviewer for several journals and conferences in his area of research. He also served as the Vice-Chair of the IEEE SMC Society, Windsor Section, from 2019 to 2022.\\
\end{IEEEbiography}

\vspace{-1.43cm}

\begin{IEEEbiography}[{\includegraphics[width=1in,height=1.25in,clip,keepaspectratio]{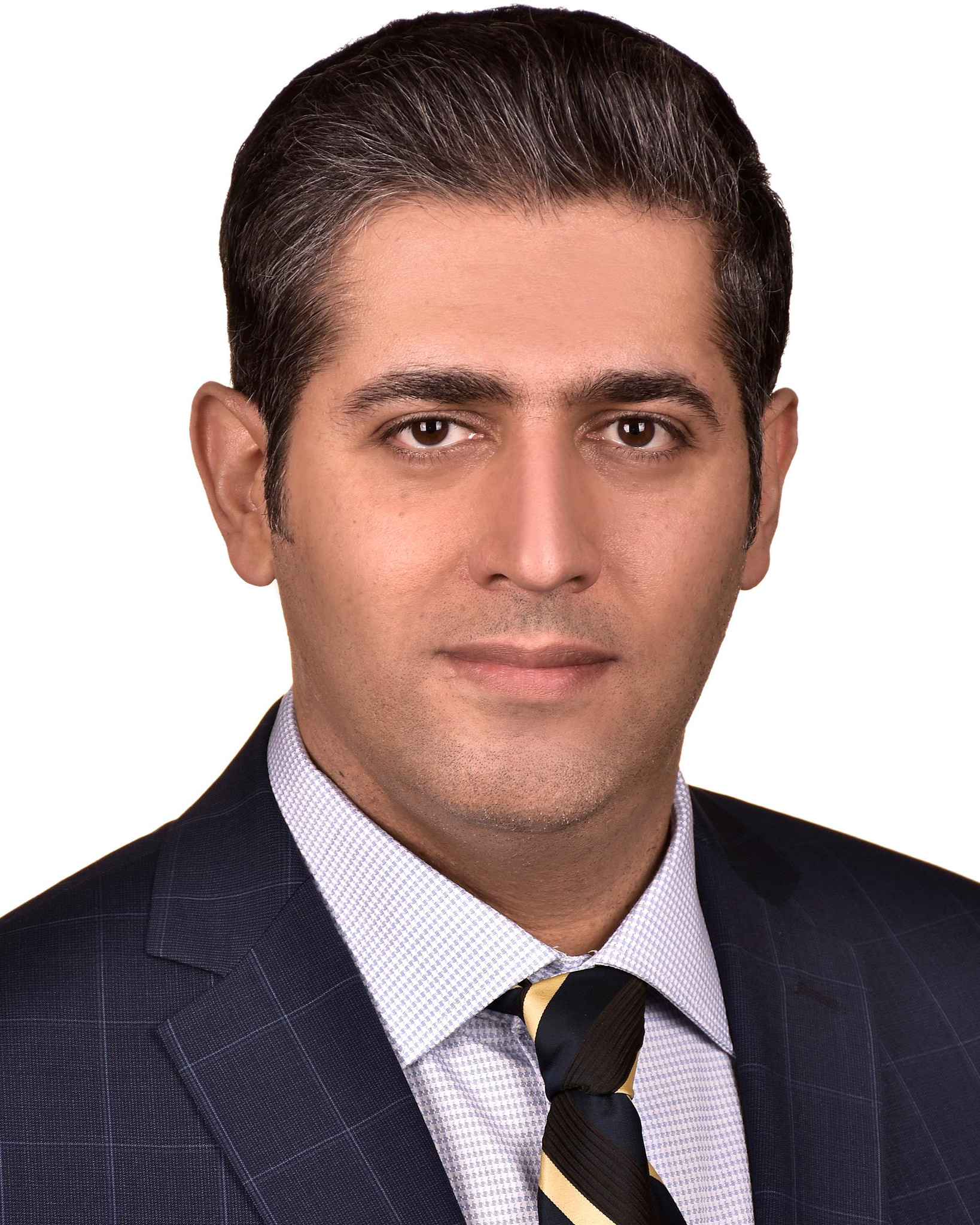}}]{Roozbeh Razavi-Far}
(Senior Member, IEEE) is an Assistant Professor at the Faculty of Computer Science and Canadian Institute for Cybersecurity, at the University of New Brunswick. His research focuses on machine learning, big data analytics, computational intelligence, and cybersecurity of cyber-physical systems. He has authored or co-authored more than 150 papers in scholarly journals and international conferences. Stanford lists his name among the top two percent most cited researchers for 2022. He is the recipient of several awards and grants including NSERC-DG, NSERC-ECR, NBIF, USRG and NSERC-PDF. He is an Associate Editor at several journals, including \textit{Neurocomputing}, \textit{Machine Learning with Applications}, \textit{Discover Artificial Intelligence}, and \textsc{IEEE Transactions on Industrial Cyber-Physical Systems}. He served as a Guest Editor and Chair for several journals and peer-reviewed conferences, and the Chapter Chair of IEEE Computational Intelligence, and Systems, Man and Cybernetics Societies at Windsor Section.\\
\end{IEEEbiography}

\vspace{-1.43cm}

\begin{IEEEbiography}[{\includegraphics[width=1in,height=1.25in,clip,keepaspectratio]{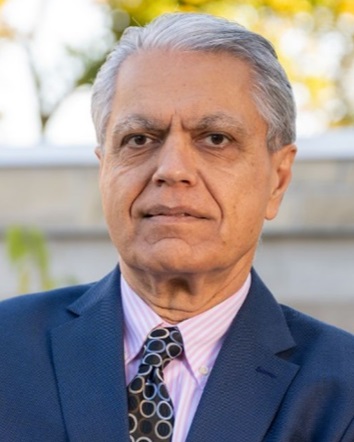}}]{Mehrdad Saif}(Fellow, IEEE) is a distinguished figure in the field of systems and control, with a career spanning nearly four decades. He received the B.S., M.S., and D.Eng. degrees in electrical engineering from Cleveland State University, OH, USA, in 1982, 1984, and 1987, respectively. Throughout his graduate studies, he was involved in research sponsored by NASA Lewis (now Glenn) Research Center and the Cleveland Advanced Manufacturing Program (CAMP). In 1987, Dr. Saif joined Simon Fraser University's School of Engineering Science as an Assistant Professor. He rose through the ranks, becoming a Full Professor in 1997. From 2002 to 2011, he took on the role of Director of the School, spearheading its significant expansion during his tenure. Subsequently, he served as the Dean of the Faculty of Engineering at the University of Windsor from July 2011 until September 2021. There, he initiated major developments, including substantial enrollment growth and the addition of new programs in aerospace engineering, engineering management, B.Eng. technology, mechatronics, among others. Under his leadership, the Faculty of Engineering at University of Windsor also saw an increase in both faculty/staff numbers and research output. Dr. Saif's research contributions are both vast and impactful, specializing in systems and control, estimation, observer theory, and AI/ML-based approaches to fault diagnostics and condition monitoring. His work has applications in a range of sectors, including automotive, power, and autonomous systems. To date, he has published over 400 refereed journal and conference papers and edited a book in these subject areas. He has garnered over 10,000 citations and holds a Google h-index of 50. Moreover, Dr. Saif is listed in Stanford University's database of the top 100,000 career scientists from 1965 to the present. In his own area of expertise, the same database ranks him in the top 0.7\%. Research.com also places him at 2681 worldwide and 139Th nationally among all electrical and electronics engineers. Dr. Saif expertise has been sought after by notable organizations such as GM, NASA, B.C. Hydro, and Canadian Space Agency (CSA). He has also significantly contributed to the IEEE Control Systems Society as the Chair of its Vancouver Section and serves on the Editorial Board of several esteemed IEEE journals such as the \textsc{IEEE Access}, \textsc{IEEE SMC Magazine}, \textsc{IEEE Industrial Informatics}, \textsc{IEEE Transactions on Industrial and Cyber-physical Systems}, among others. Dr. Saif is a Registered Professional Engineer in Ontario, Canada, and has been honored with Fellowships in several prestigious organizations, including IEEE, the Canadian Academy of Engineering (CAE), the Engineering Institute of Canada (EIC), the Institution of Engineering and Technology (IET), and Asia-Pacific Artificial Intelligence Association (AAIA). His IEEE fellowship is particularly noteworthy, earned \textit{“for contributions to monitoring, diagnosis and prognosis in cyber-physical health systems”}.
\end{IEEEbiography}

\begin{IEEEbiography}[{\includegraphics[width=1in,height=1.25in,clip,keepaspectratio]{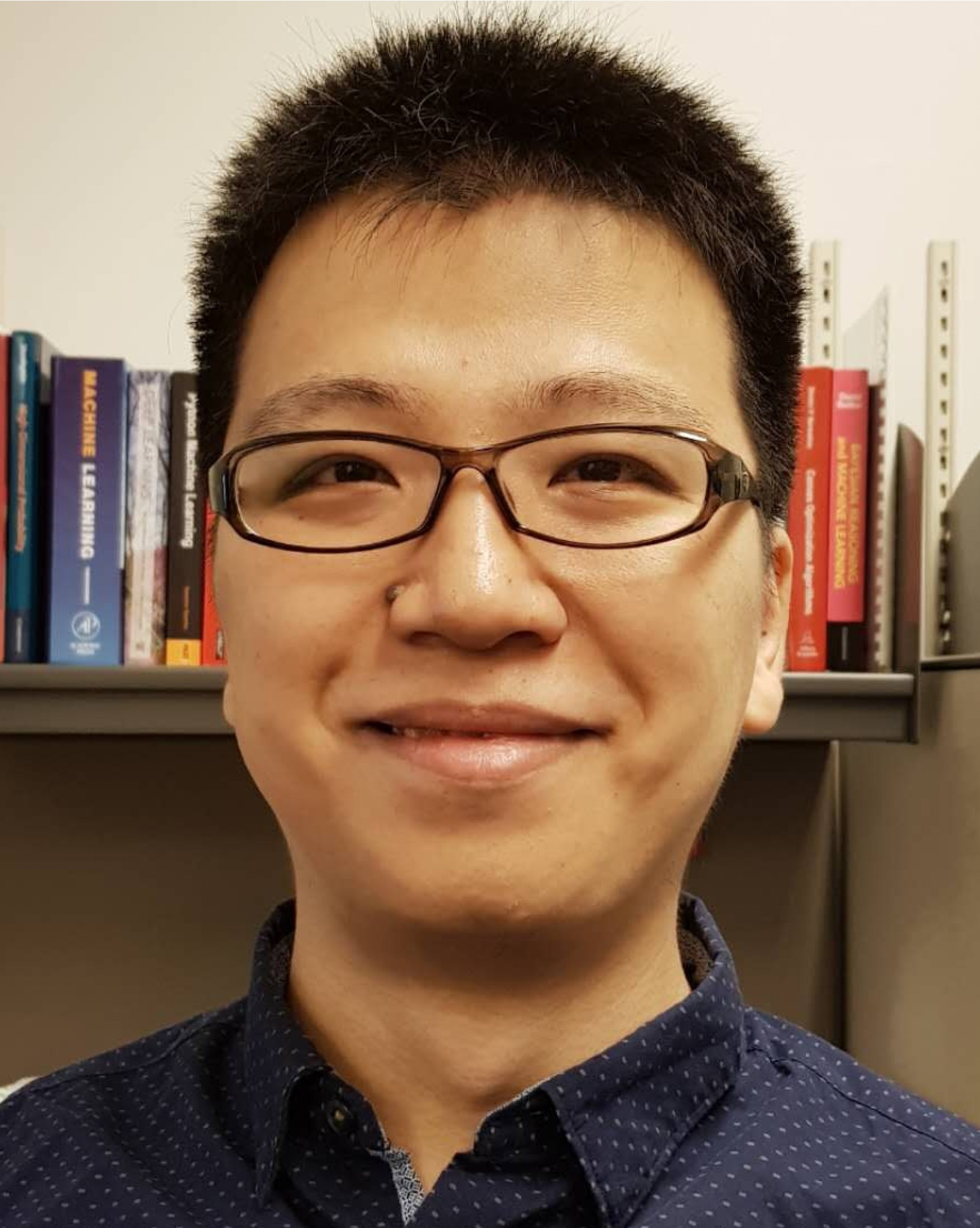}}]{Boyu Wang} (Member, IEEE) 
received his B.Eng. degree in Electronic Information Engineering from Tianjin University, Tianjin, China, M.Sc. degree in Electrical and Computer Engineering from University of Macau, Macau, China, and Ph.D. in Computer Science from McGill University, Montreal, QC, Canada. He is currently an Assistant Professor with the Department of Computer Science, University of Western Ontario, London, ON, Canada. He is also affiliated with the Brain and Mind Institute and the Vector Institute. He was a Post-Doctoral Research Fellow at the University of Pennsylvania and Princeton University. His research interests include machine learning theory, algorithms, and applications.\\
\end{IEEEbiography}

\begin{IEEEbiography}
[{\includegraphics[width=1in,height=1.25in,clip,keepaspectratio]{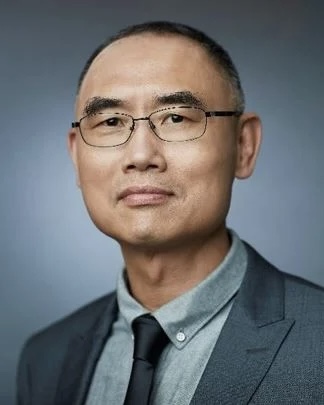}}]{Qiang Yang}
(Fellow, IEEE) received the B.Sc. degree in astrophysics from Peking University, Beijing, China, in 1982, and the M.Sc. degree in astrophysics and the Ph.D. degree in computer science from the University of Maryland, College Park, MD, USA, in 1985 and 1989, respectively. He was a Faculty Member with the University of Waterloo, Waterloo, ON, Canada, from 1989 to 1995, and Simon Fraser University, Burnaby, BC, Canada, from 1995 to 2001. He was the Founding Director of Huawei’s Noah’s Ark Lab, Hong Kong, from 2012 to 2014 and a Co-Founder of 4Paradigm Corporation, Beijing, an artificial intelligence (AI) platform company. He is currently the Head of the AI Department and the Chief AI Officer of WeBank, Shenzhen, China. Dr. Yang has been a Professor Emeritus with the Computer Science and Engineering Department, Hong Kong University of Science and Technology (HKUST), Hong Kong, since 2023, where he was previously a Chair Professor, and a former Head of the CSE Department and the Founding Director of the Big Data Institute from 2015 to 2018. He is the author of several books, including \textit{Intelligent Planning} (Springer), \textit{Crafting Your Research Future} (Morgan \& Claypool), and \textit{Constraint-Based Design Recovery for Software Engineering} (Springer). His research interests include AI, machine learning, and data mining, especially in transfer learning, automated planning, federated learning, and case-based reasoning. Dr. Yang has served as an Executive Council Member of the Advancement of AI (AAAI) from 2016 to 2020. He is a fellow of several international societies, including ACM, AAAI, IEEE, IAPR, and AAAS. He was a recipient of several awards, including the 2004/2005 ACM KDDCUP Championship, the ACM SIGKDD Distinguished Service Award in 2017, and the AAAI Innovative AI Applications Awards in 2018 and 2020. He was the Founding Editorin-Chief of the \textit{ACM Transactions on Intelligent Systems and Technology} and \textsc{IEEE Transactions on Big Data}. He has served as the President of the International Joint Conference on AI (IJCAI) from 2017 to 2019.\\
\end{IEEEbiography}

\end{document}